\documentclass[graybox, nosecnum]{svmult}

\usepackage{graphicx}        % standard LaTeX graphics tool
\usepackage{amsmath}                             % when including figure files
\usepackage[square,numbers]{natbib}

\def\prl{PRL}
\def\prd{PRD}
\def\apj{ApJ}
\def\apjl{ApJL}
\def\apjs{ApJS}

\def\aap{A\&A}
\def\mnras{MNRAS}
\def\araa{ARA\&A}

\def\nat{Nature}
\def\aj{AJ}

\begin{document}
\title*{Gamma-Ray Bursts}
\author{Yun-Wei Yu \thanks{corresponding author}, He Gao, Fa-Yin Wang, and Bin-Bin Zhang}
\institute{Yun-Wei Yu \at Institute of Astrophysics, Central China Normal University, Wuhan 430079, China\\ \email{yuyw@ccnu.edu.cn}
\and He Gao \at Department of Astronomy, Beijing Normal University, Beijing 100875, China\\ \email{gaohe@bnu.edu.cn}
\and Fa-Yin Wang \at Key Laboratory of Modern Astronomy and Astrophysics, Ministry of Education, China
\at School of Astronomy and Space Science, Nanjing University, Nanjing 210093, China
 \\ \email{fayinwang@nju.edu.cn}
\and Bin-Bin Zhang
\at Key Laboratory of Modern Astronomy and Astrophysics, Ministry of Education, China
\at School of Astronomy and Space Science, Nanjing University, Nanjing 210093, China
 \\ \email{bbzhang@nju.edu.cn}}

\maketitle
\abstract{ Gamma-ray bursts (GRBs) are short and intense bursts of $\sim$100 keV$-$1MeV photons, usually followed by long-lasting decaying afterglow emission in a wide range of electromagnetic wavelengths from radio to X-ray and, sometimes, even to GeV gamma-rays. These emissions are believed to originate from a relativistic jet, which is driven due to the collapse of special massive stars and the mergers of compact binaries (i.e., double neutron stars or a neutron star and a black hole). This chapter first briefly introduces the basic observational facts of the GRB phenomena, including the prompt emission, afterglow emission, and host galaxies. Secondly, a general theoretical understanding of the GRB phenomena is described based on a relativistic jet's overall dynamical evolution, including the acceleration, propagation, internal dissipation, and deceleration phases. Here a long-lasting central engine of the GRBs can substantially influence the dynamical evolution of the jet. In addition, a supernova/kilonova emission can appear in the optical afterglow of some nearby GRBs, which can provide an important probe to the nature of the GRB progenitors. Finally, as luminous cosmological phenomena, it is expected to use GRBs to probe the early universe and to constrain the cosmological parameters.}

\section{Keywords}
gamma-ray burst; afterglow; supernova; kilonova; transient; relativistic jet; shock; radiation mechanism; gravitational wave; cosmology

\section{Introduction}

Gamma-ray bursts (GRBs) are short and intense bursts of $\sim$100 keV$-$1MeV photons, which were first discovered in the year of 1967 by the Vela satellites and announced to the public in 1973 as astronomical phenomena \citep{Klebesadel1973}. The BATSE detector onboard the COMPTON Gamma-Ray Observatory, which was launched in 1991, revealed the isotropic distribution of GRBs on the sky \cite{Meegan1992}. Such a distribution indicated the GRBs have an extra-galactic origin at cosmological distances rather than originate from our Galaxy, which was further confirmed in 1997 by the measurements of the cosmological redshifts of several GRBs. In that year, the new-launched Italian/Dutch BeppoSAX satellite captured an X-ray counterpart of GRB 970228 \cite{Costa1997} and the X-ray emission can exceed the sensitivity of the detector for about one week. Thanks to the X-ray observation, the accuracy of the GRB localization was enhanced significantly, which made it possible to monitor the transient counterparts also in the optical and radio bands \cite{van Paradijs1997,Frail1997} and as well as to identify the host galaxy. Therefore, the redshift of the GRB can be obtained from the absorption or emission lines. These long-lasting multi-wavelength transient counterparts of GRBs are now well known as the afterglow emission, while the sub-MeV gamma-ray emission detected by the GRB triggering detectors is termed as the prompt emission.

The cosmological origin of GRBs indicates them to be the most explosive phenomena in the universe after the Big Bang. Meanwhile, the rapid variability of the GRB prompt emission gives a stringent constraint on the size of the emission source. Therefore, it is believed that GRBs should be produced by a relativistic jet launched by a catastrophic event of a stellar system, including the collapse of a special category of massive stars \cite{1993ApJ...405..273W} and the merger of compact binaries \cite{Paczynski1986,Eichler1989}. The former hypothesis was quickly confirmed by the discovery of a supernova emission emerging from the optical afterglow of some nearby GRBs, as represented by the GRB 980425/SN 1998bw, GRB 030329/SN 2003dh, and GRB 060218/SN 2006aj association events. On the contrary, the merger origin hypothesis is long pending until the observation of GRB 170817A which was confidently associated with a gravitational wave event GW170817 \cite{Abbott2017a,Abbott2017b}. Then, it becomes the crucial task of the GRB research to answer how these stellar catastrophic events can generate a relativistic jet and how such a jet produces the GRB prompt and afterglow emission. Additionally, as one of the most distant detectable objects, it is also highly concerned whether and how we can use the GRBs as a tool to probe the universe.

\section{Observations}
\subsection{Prompt Emission}

\begin{figure}
\begin{center}

	\includegraphics[width=0.95\textwidth]{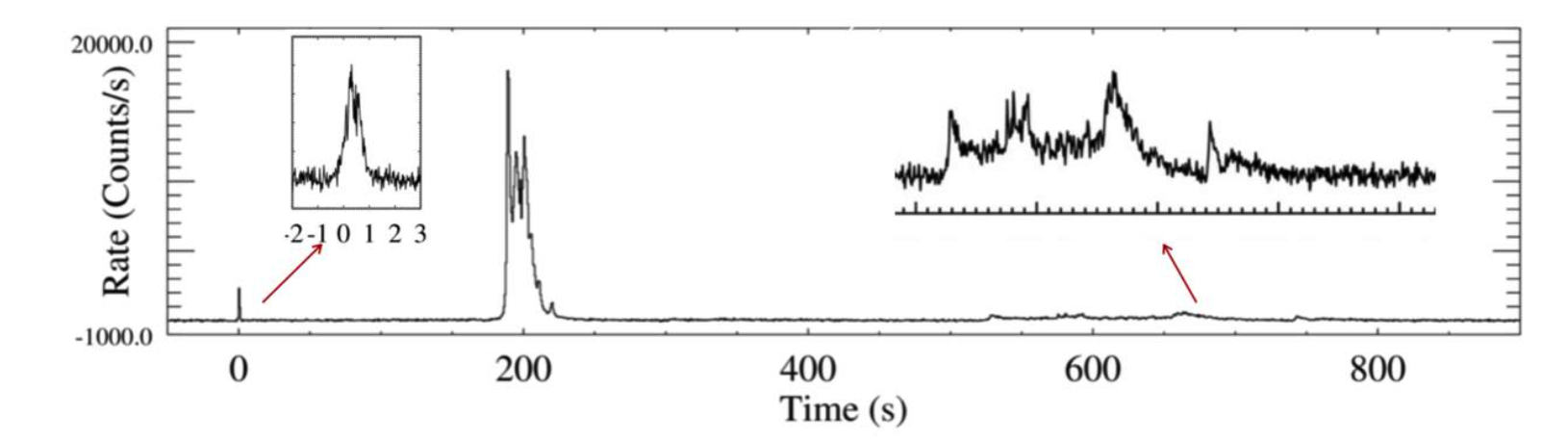}
	\caption{The light curve of GRB 160625B, which illustrate a precursor, a main burst and an extended emission phase. Adapted from\cite{2018NatAs...2...69Z}.
	\label{GRBlc}}
	\end{center}

\end{figure}

The prompt emission of GRBs is represented by their complex, highly variable, and unpredictably shaped pulses, which constitute their light curves in a typical energy range between the X-ray and $\gamma$-ray bands. The prompt emission of a GRB may consist of several consecutive episodes separated by quiet periods in-between. In addition to the main burst, some, if not all, GRBs are presumed to have a pre-burst phase (``precursor") and a post-burst phase (``extended emission"), as shown in Fig. \ref{GRBlc}. The duration of GRB prompt emission is usually defined by the so-called $T_{90}$, i.e., the time interval in which the integrated photon counts increase from 5\% to 95\% of the total counts. Based on the bimodal distribution of GRBs in the duration-hardness diagram (the hardness of a burst is usually denoted as the photon count ratio in two fixed observational energy bands), GRBs were classified into two categories: the long-duration, soft-spectrum class (LGRBs) and the short-duration, hard-spectrum class (SGRBs) \citep{1993ApJ...413L.101K}. Since both $T_{90}$ and hardness ratio are energy- and sensitivity-dependent, the separation line between the long and short GRBs remains unclear but is conventionally set at about 2s. Such an empirical dichotomy could be a natural result of the two possible different origins mentioned above, i.e., the core collapse of Wolf$-$Rayet stars for long GRBs and the mergers of two compact objects for short GRBs.

% The prompt gamma-ray emission is usually highly variable,

The temporal properties of prompt emission vary by a burst-by-burst basis. Their light curve shapes vary from a single smooth pulse to extremely complex light curves with many erratic pulses with different durations, amplitudes, and fine structures \citep{1995ARA&A..33..415F}. The power density spectra analysis shows no indication that the light curve is periodic \citep{2000ApJ...535..158B}. The light curves may be decomposed as the superposition of an underlying slow component and a more rapid fast component \citep{2012ApJ...748..134G}. Moreover, the light curves would vary with energy. The fast component tends to be more significant in high energies and becomes less significant at lower frequencies \citep{2006A&A...447..499V}. The shape of the slow component is typically asymmetric, usually with a fast-rising exponential-decay (FRED) shape \citep{1996ApJ...459..393N}. In more detail, the slow component pulse tends to be narrower in harder energy bands and wider in softer energy bands \citep{2005ApJ...627..324N}. The arrival time of a pulse in a softer band is typically delayed (or ``lagged") with respect to the arrival time in a harder band, which is called the ``spectral lag" effect \citep{2000ApJ...534..248N}. %Additionally, a fraction of GRBs have a precursor emission component well separated from the main burst, which is typically softer than the main burst \citep{2009A&A...505..569B}.

GRB spectra are typically non-thermal. When the detector's energy band is wide enough, the photon number spectrum of prompt emission, both for time-resolved spectrum and time-integrated spectrum, can usually be fitted with a broken power law known as the Band function in form of \citep{Band1993}
\begin{equation}
N(E) =\left\{ \begin{array}{ll}
A\left(\frac{E}{100\rm{keV}}\right)^\alpha {\rm e}^{-E/E_0} &, E<(\alpha-\beta)E_0\\
A\left[\frac{(\alpha-\beta)E_0}{100\rm{keV}}\right]^{\alpha-\beta} {\rm e}^{\beta-\alpha} (\frac{E}{100\rm{keV}})^\beta &, E \ge (\alpha-\beta)E_0
\end{array}\right.\label{eq:band function}
\end{equation}
where $A$ is the normalization factor, $E_0$ is the break energy in the spectrum, $\alpha$ and $\beta$ are the low-energy and high-energy photon spectral indices with $\alpha$ ranging at $(-2, 0)$ and $\beta$ ranging at $(-4, -1)$ \citep{2000ApJS..126...19P}. The peak energy of the $E^2N(E)$ spectrum, $E_{\rm p} = (2+\alpha) E_0$, distributes within several orders of magnitude but clusters around $200-300 \rm{keV}$ \citep{2000ApJS..126...19P,2013ApJS..208...21G}. When the detector's energy band is not wide enough or a GRB is not bright enough, the photon number spectrum of the prompt emission can behave as a cut-off power-law, which reads
\begin{equation}
 N(E) =
 A\left(\frac{E}{100\rm{keV}}\right)^\alpha {\rm e}^{-E/E_{\rm c}},
 \label{eq:band function}
\end{equation}
where $E_{\rm c}$ is the cutoff energy and the peak energy in the $E^2N(E)$ spectrum of this model is $E_{\rm p} = (2+\alpha) E_c$. The spectrum might even behave as a simple power law as
\begin{equation}
 N(E) = A\left(\frac{E}{100\rm{keV}}\right)^{-\hat{\Gamma}},
 \label{eq:band function}
\end{equation}
where $\hat{\Gamma}$ is the photon index, which is positive by definition.

\begin{figure}
\begin{center}

	\includegraphics[width=0.8\textwidth]{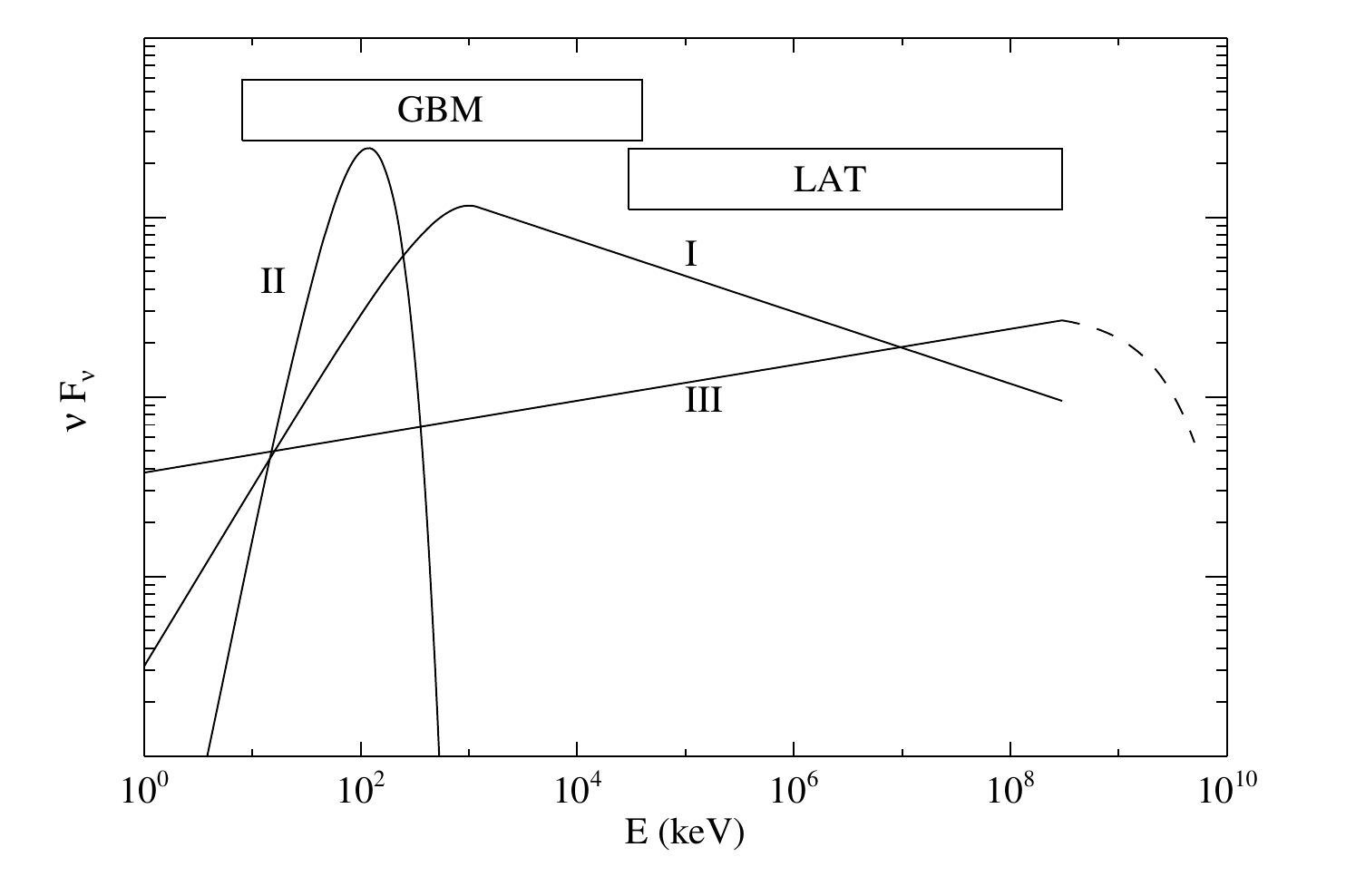}
	\caption{A illustration of the three possible spectral components of GRB prompt emission. From \cite{2011ApJ...730..141Z}}\label{GRBspectrum}
	\end{center}

\end{figure}

Although the main component of GRB spectra is non-thermal and ``Band"-like, a few GRBs, however, show that an additional (quasi-)thermal component and/or high-energy power-law component is required to explain the spectral data \citep{2011ApJ...730..141Z}. For instance, in the spectra of several Fermi bursts (e.g., GRBs 100724B,  110721A, and  120323A), a sub-dominant thermal component appears at the left shoulder of the Band component. Additionally, the superposition between the Band and high-energy components has been seen in GRB 090926A and other GRBs. In the exceptional case of GRB 090902B, the spectra only contain the superposition between a thermal component and a high-energy spectral component. Detailed analysis of the time-resolved spectra of bright bursts shows that significant spectral evolution is usually observed between $E_{\rm p}$ and the flux with two typical types of evolution patterns \citep{2012ApJ...756..112L}, namely the hard-to-soft and intensity-tracking patterns. A sketch illustrating the combinations of the three elemental spectral components of prompt emission is shown in Fig. \ref{GRBspectrum}.

Finally, it is worth pointing out that broadband emission is naturally expected and had been indeed observed during the prompt phase. Such observations include prompt X-ray flares \citep{2014ApJ...789..145H}, prompt optical flashes \citep{1999Natur.398..400A}, and prompt GeV flashes \citep{2014Sci...343...42A}.  Different patterns are observed for the relationship between the prompt sub-MeV emission and those in other wavelengths. For instance, both mismatch and tracking behaviors exist between the optical and gamma-ray peaks, and bright X-ray emission was observed during the quiescent phase of sub-MeV emission. Furthermore, delayed onset of the GeV emission is usually observed with respect to the MeV emission. Therefore,  multiple emission sites are usually required to account for these different behaviors of the prompt emission components.

\subsection{Afterglow and Associated Supernova/Kilonova}
% In the early years of GRB afterglow observations, the data can be obtained always at a relatively late stage, where the light curve in each band generally displays a power-law decay behavior. Such a situation had finally been changed by the launch of the NASA's Swift mission. Thanks mainly to the rapid slewing and precise localization capability of the X-Ray Telescope (XRT) onboard the Swift, unprecedented new information about GRB afterglows was revealed, especially in the early phases.

Because of the limited observational facilities, early studies of GRB afterglows have relied primarily upon late-time data, which are typically obtained hours or days after the burst. A power-law decay can be observed at all wavelengths, as expected by the standard forward shock model. Such a picture was altered by the launch of NASA Swift mission, which brings unprecedented early X-ray data of GRB afterglow from its X-Ray Telescope (XRT) thanks to its rapid slewing and precise localization capability.

\begin{figure}
\begin{center}

	\includegraphics[width=0.8\textwidth]{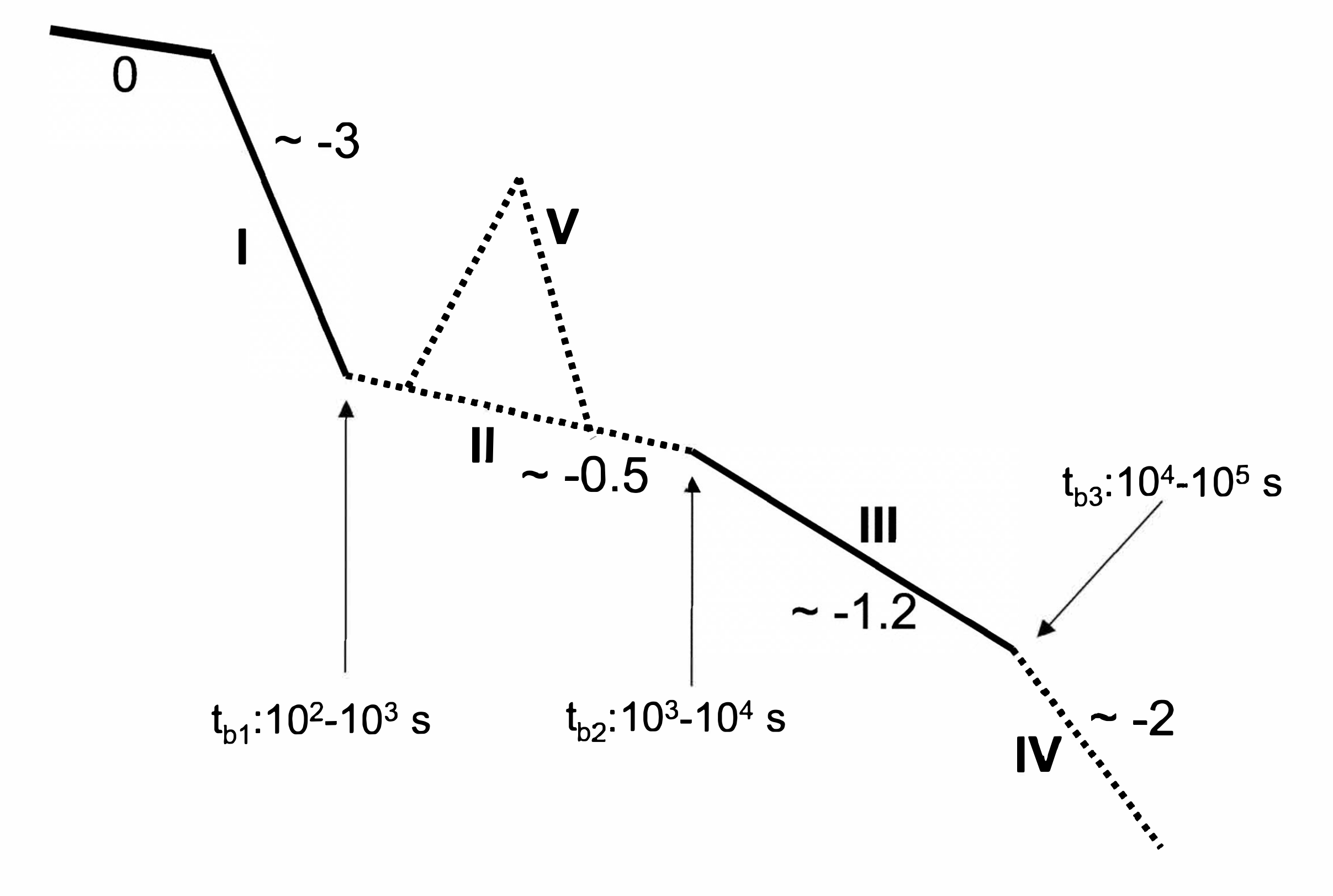}
	\caption{A cartoon illustration of the X-ray afterglow light curve based on the observations of the Swift XRT. From \cite{2006ApJ...642..354Z}}\label{Xrayag}.
	\end{center}
\end{figure}

As shown in Fig. \ref{Xrayag}, the X-ray afterglow light curves can be summarized as a canonical pattern composed of five components, namely a distinct rapidly decaying component, a shallow decay component, a normal decay component, a post jet break component, and X-ray flares \citep{2006ApJ...642..389N,2006ApJ...642..354Z}. Each of those components is further itemized below.

\begin{itemize}
 \item The rapidly decaying component (I) behaves as the ``tail" of the prompt emission, whose decay slope is typically in the range of $\sim3$ to $\sim10$ \citep{2005Natur.436..985T}. A good fraction of GRBs showed a clear hard-to-soft evolution during the rapid decay \citep{2007ApJ...666.1002Z}.
 \item The shallow decay component (II) is usually adjacent to the steep decay phase and is followed by a normal decay. The typical slope of a shallow decay component is in the range between $0$ and $\sim0.7$. There is no significant spectral evolution between the shallow decay segment and its follow-up segment, and the latter is usually consistent with the external-shock models \citep{2007ApJ...670..565L,2019ApJ...883...97Z}. If the slope of the shallow decay component is close to 0, it is also called a plateau. In rare cases, an X-ray plateau can be followed by a very steep decay with a decay slope steeper than $-3$.  \citep{2007ApJ...665..599T}, which is called internal plateau because it probably arises from an internal dissipation process rather than the external shock.
 \item A normal decay component (III) usually follows the shallow decay component, or sometimes directly follows the steep decay component. It has a decay slope of $\sim1$, which is the typical value predicted in the standard external forward shock model.
 \item A late steeper decay (with a typical slope of approximately $-2$) component (IV) often follows the normal decay segment, which is also expected in the external forward shock model due to the so-called jet break effect \citep{1998ApJ...503..314P}.
 \item X-ray flares (V) are considered to have different origins from other components since they are ``superposed" on those background power-law decay components \citep{2005Sci...309.1833B,2007ApJ...671.1903C}. Their light curves typically show rapid rise and fall with steep indices. Many properties of X-ray flares in both temporal and spectral domains are analogous to prompt emission, suggesting that they might be directly powered by the GRB central engine, similar to prompt emission.

\end{itemize}

Compared with the X-ray afterglow, the optical afterglow exhibits more complicated behavior. An optical afterglow light curve can also consist of shallow decay, normal decay, and jet-break phases. Additionally, optical flares were also observed in certain instances. The properties of these components are similar to those observed in the X-ray band. In late stages, a re-brightening feature is occasionally observed  in the optical light curve of some GRB, which could be related to the jet structure \citep{2011A&A...531A..39N}. One point worthy of note is that a large proportion of GRBs have no detectable optical afterglow. This is primarily due to the heavy dust extinction within the GRB host galaxies.

\begin{figure}
\begin{center}
	\includegraphics[width=0.7\textwidth]{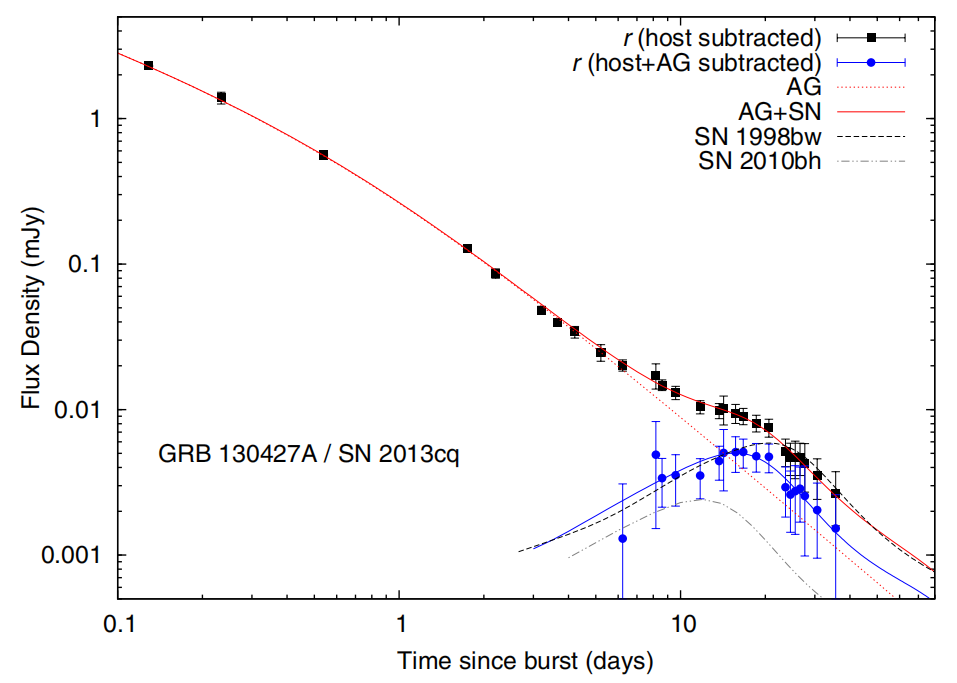}
	\caption{The $R-$band light curve of GRB 130427A/ SN 2013cq. From \cite{2013ApJ...776...98X}}\label{SN2013cq}
\end{center}
\end{figure}

In some optical afterglows of long GRBs, a bump feature, usually with a red color, can appear about a week after the GRB trigger, as shown in Fig. \ref{SN2013cq}. Such a bump feature is usually interpreted as the signature of an associated supernova \citep{2003Natur.423..847H}. Comparatively to long GRBs, a fainter-than-supernova optical/IR transient has been predicted to accompany short GRBs that originate from neutron star-neutron star (NS-NS) or neutron star-black hole (NS-BH) mergers \citep{Li1998,Metzger2010}. Such a transient is now commonly referred to as a kilonova. The kilonova prediction had been first confirmed by the discovery of the optical transient AT2017gfo following the GW170817/GRB 170817A event \cite{Arcavi2017,2017Sci...358.1570D}. Additionally, several kilonova candidates were also claimed in several short GRB events, including GRB 130603B and GRB 050709 \citep{2009Natur.461.1254T,2015NatCo...6.7323Y,2015ApJ...807..163G}.

% In the optical afterglows of some long GRBs, a bump feature, usually with a red color, shows up about a week after the GRB trigger, as shown in Fig. \ref{SN2003dh}. This signature is usually interpreted as the signature of an associated supernova \citep{2003Natur.423..847H}. Different from the long GRBs, a fainter-than-supernova optical/IR transient has been predicted to be associated with neutron star-neutron star (NS-NS) or neutron star-black hole (NS-BH) mergers \citep{Li1998,Metzger2010}, which is now widely termed as kilonova. Such a kilonova prediction had been first confirmed by the discovery of the optical transient AT2017gfo following the GW170817/GRB 170817A event \cite{Arcavi2017,2017Sci...358.1570D}, as shown in Fig. \ref{AT2017gfo}. Additionally, several kilonova candidates had also been claimed to be found from the afterglow emission of some pre-GRB 170817A short GRBs such as GRB 130603B \citep{2009Natur.461.1254T,2015NatCo...6.7323Y,2015ApJ...807..163G}.

% In comparison, the radio and high-energy afterglows show relatively simple behaviors. The statistical studies show that the radio afterglow usually simply rises and reaches a peak around 3-6 days, which is well consistent with the standard external forward shock model. At high energies, it is not as easy to identify afterglow components as in the low energy bands, because the number of photons greatly reduces. The high-energy afterglow typically shows a power law decay with time, which could also be a result of the external forward shock emission, either with a synchrotron radiation or related to a synchrotron self-Compton (SSC) spectral component.

The radio and high-energy afterglows, in contrast, exhibit relatively simple behaviors. According to statistical studies, the radio afterglow typically rises and reaches a peak around 3-6 days. This is consistent with the standard external shock model. The detection of afterglow components at high energies is more challenging than at low energies as the number of photons dramatically drops at higher energies. The high-energy afterglow typically shows a power-law decay with time, which could result from external forward shock emission, either with synchrotron radiation or related to a synchrotron self-Compton (SSC) spectral component.

\subsection{Host Galaxy}

For long GRBs, most host galaxies are irregular, star-forming galaxies, with a few being spiral galaxies with active star formation \citep{2006Natur.441..463F}. Nevertheless, it is suggested that long GRB hosts are relatively metal-poor compared with field galaxies. Within the host galaxy, most long GRBs reside in the brightest core regions, where the specific star formation rate is the highest \citep{2008AJ....135.1136M}. These facts appear to support the massive star origin of long GRBs.
Unlike long GRBs, the majority of host galaxies of short GRBs are elliptical or early type \citep{2005Natur.437..851G}.
Some short GRBs are even host-less, which may be ``kicked" away from their host. Moreover, most short GRBs are found to be far from the bright light of the host galaxies \citep{2010ApJ...708....9F}.
These discoveries indicate that the short GRBs are likely not directly associated with the deaths of massive stars but are more consistent with the compact binary mergers.

%---------------------------------------------------------------------------------------------------------------%
%---------------------------------------------------------------------------------------------------------------%
%---------------------------------------------------------------------------------------------------------------%
%---------------------------------------------------------------------------------------------------------------%
\section{Theory}
\subsection{Central Engine and Jet}\label{sec:intro}
\subsubsection{Energy Sources}

Assuming the GRB emission originates from a relativistic jet of a Lorentz factor $\Gamma$, the size of the emission region of GRBs can be constrained at the scale of
\begin{equation}\label{grbradius}
R_{\rm GRB}=2\Gamma^2c\delta t\sim 6\times10^{12}\Gamma_{2.5}^2~\delta t_{-3}{\rm cm},
\end{equation}
where $\delta t$ is the variability timescale of the GRB light curves. Hereafter the conventional notation $Q_x=Q/10^x$ is used in cgs units. Eq. (\ref{grbradius}) indicates the progenitors of GRBs can only be in the order of stellar systems. Meanwhile, for the typical fluence $F\sim(10^{-7}-10^{-4})\rm  erg~cm^{-2}$ of GRBs and their cosmological distances, the isotropically-equivalent energy release of these GRBs can be calculated as
\begin{equation}
E_{\rm iso}=4\pi d_{\rm L}^2F/(1+z)\sim(1+z)^{-1}10^{50-53}~\rm erg,
\end{equation}
where $d_{\rm L}\sim10^{28}\rm cm$ is a typical luminosity distance and $z$ is redshift. Even considering that the GRB emission is actually produced by a highly-beamed jet, the realistic energy release can still be as high as $\sim10^{50-51}$ erg. For a stellar system, such a huge energy release can only be supplied by the gravitational energy from the catastrophical contraction of the system from a large size to several to a few tens of kilometers:
\begin{equation}
\Delta U\sim {3\over5}{GM^2\over R}\sim 10^{53}\left({M\over M_{\odot}}\right)^{2}\left({R\over 10{\rm km}}\right)~\rm erg,
\end{equation}
where $M$ is the mass of the system and $R$ is the radius after the system contraction. Specifically, such a contracting stellar system could be a collapsing massive star or a merging compact binary, which corresponds to the observational long and short GRBs, respectively.

Following the above considerations, the central engine of GRBs should be a compact object (BH or NS) forming from the stellar collapses and compact binary mergers. How such a compact object launches relativistic jets and extracts the released gravitational energy is still an open question. The most promising mechanism is the hypercritical accretion onto the central engine during its forming, where the typical accretion rate, i.e., $\dot{M}\sim (0.1-1)M_{\odot}~s^{-1}$, can in principle be high enough to explain the GRB luminosities:
\begin{equation}
L_{\rm GRB}=\zeta \dot{M}c^2=2\times10^{51}\zeta_{-3}\left({\dot{M}\over 1\rm M_{\odot}~s^{-1}}\right),
\end{equation}
where $\zeta$ is the radiation efficiency of the accretion. About a half of the released gravitational energy is initially stored as the internal energy of the accretion disk, which leads the disk to be so hot that a great abundance of neutrinos/antineutrinos are produced. These neutrinos/antineutrinos can bring away the internal energy easily because they can escape from the transparent disk freely. The $\nu\bar{\nu}$ luminosity can be obtained by calculating the structure and ingredients of the neutrino-dominated accretion flow (NDAF) in the disk \citep{Qian1996}.

Besides the internal energy, the gravitational energy can also be converted into the rotational energy of the central engine and the disk. If the central engine is a BH, its rotational energy can be expressed as
\begin{equation}
E_{\rm rot}=2\times10^{54}f(a_*){M_{\rm bh}\over M_{\odot}}\rm erg,
\end{equation}
where $f(a_*)=1-\sqrt{(1+q)/2}$, $q=\sqrt{1-a_*^2}$, and $a_*=Jc/GM_{\rm bh}^2$ is the dimensionless spin parameter with $J$ and $M_{\rm bh}$ being the angular momentum and mass of the BH. For $a_*=0.1$, we have $f(a_*)\approx 0.001$ and $E_{\rm rot}\approx10^{51}\rm erg$. When this rapidly spinning BH is connected with a remote astrophysical load by magnetic field
lines, its rotational energy can be extracted through the Blandford-Znajek (BZ) mechanism \citep{Blandford1977}. On the other hand, if the central engine is an NS, then its rotational energy can reach
\begin{equation}
E_{\rm rot}={1\over2}I\Omega^2=2\times10^{52}I_{45}P^{-2}_{-3}\rm erg,
\end{equation}
where $I$ is the moment of inertial of the NS, $\Omega$ and $P$ are the spin frequency and period. In this case, if the dipolar magnetic field of the NS can be as high as $B_{\rm p}\sim 10^{16}$ G, then its rotational energy can be released in a timescale of $t_{\rm sd}=20I_{45}R_{\rm s,6}^{-6}B_{\rm p,16}^{-2}P_{-3}^2$ s, which could also play a role in driving GRB jet, where $R$ is the radius of the NS. In these cases, the jet could be dominated by a Poynting flux.% rather t $\nu\bar{\nu}$ outflow

\subsubsection{Jet Acceleration}
Because of the extremely high density of the released energy, neutrinos and anti-neutrinos coming from the hypercritical accretion disk can annihilate above the disk to produce photons and electron-positron pairs, which are highly coupled with each other to form a fireball \citep{Paczynski1986,Meszaros1993}. For a fireball in thermal equilibrium, its electron-positron density can be calculated as
\begin{equation}
 n_{\pm}=\frac{1}{2}\left(\frac{2 m_{\rm e} k_{\rm B} T}{\pi \hbar^{2}}\right)^{3 / 2} \exp \left(-\frac{m_{\rm e} c^{2}}{k_{\rm B} T}\right),
\end{equation}
where $T$ is the temperature of the fireball satisfying ${k}_{\rm B} {T} \ll m_{\rm e} {c}^{2}$. The fireball can expand drastically due to the high radiation pressure. Then, a certain mass of baryons can be stripped by the fireball from the disk to be coupled with the pairs.

The dynamical evolution of the expanding fireball can be described by relativistic hydrodynamics, which shows the expanding fireball is self-similar and can be approximated by a thin shell \citep{Meszaros1993}. Then, the conservation of the mass and energy of the shell can be written as
\begin{eqnarray}\label{fbdyn}
\Gamma r^2n'&=&{\rm const.},\\
\Gamma r^2{e'}^{3/4}&=&{\rm const.},
\end{eqnarray}
where $\Gamma$ is the Lorentz factor of the shell, and $r$ is the radius relative to the center of the system, and $n'$ and $e'$ are the densities of the particle number and energy, respectively, which are measured in the comoving frame of the fireball represented by the prime. In the radiation-dominated phase ($e'\gg n'm_{\rm p}c^2$), Eq. (\ref{fbdyn}) gives
\begin{eqnarray}\label{fbacc}
\Gamma\propto r,~n'\propto r^{-3},~e'\propto r^{-4},~T\propto \Gamma
e'^{1/4}\sim\rm const,
\end{eqnarray}
where the Lorentz factor increases linearly with the radius. Subsequently, when the fireball transits into the matter-dominated phase ($e'\ll n'm_pc^2$), we have
\begin{eqnarray}
\Gamma\sim {\rm const.},~n'\propto r^{-2},~e'\propto
r^{-8/3},~T\propto \Gamma e'^{1/4}\propto r^{-2/3},
\end{eqnarray}
where the Lorentz factor keeps constant, which is called a coasting phase.
%The overall dynamical evolution of a fireball is shown in Fig. \ref{overalldyn}, where the deceleration phases are also included.

%\begin{figure*}
%\begin{center}
%	\includegraphics[width=0.5\textwidth]{glob_n_2.pdf}	\includegraphics[width=0.5\textwidth]{glob_r_2.pdf}
%	\caption{The overall dynamical evolution of a fireball including the adiabatic acceleration, coasting, and external reverse and forward shock phases. The left and right panels correspond to a Newtonian and a relativistic reverse shock, respectively. The thick solid, thin solid, dotted, and dashed dotted lines give the average value of the Lorentz factor, the value at the forward shock, and the maximal value, and an analytic estimate. From \cite{1999ApJ...513..669K}}\label{overalldyn}
%	\end{center}
%\end{figure*}

As mentioned above, the outflow driven by the GRB engine can alternatively be Poynting-flux-dominated (PFD). In this case, the outflow would be accelerated due to the magnetic reconnection, which converts the magnetic energy to internal energy and subsequently to kinetic energy. It is not easy to describe the reconnection processes. An approximative and convenient treatment is to assume the comoving reconnection speed to be proportional to the Alfven speed $v'_{\rm A}=\sqrt{\sigma/(1+\sigma)}$ \citep{Drenkhahn2002}, where the parameter $\sigma$ represents the magnetization of the outflow. Then, for stripped
toroidal magnetic fields of a width $\lambda\sim cP$, the reconnection timescale can be estimated by $t'_{\rm r}=\lambda'/\epsilon v'_{\rm A}\propto \Gamma$ for $\sigma\gg1$. The ratio of $t'_{\rm r}$ to the dynamical timescale $t'_{\rm dyn}= r/\Gamma c$ gives the fraction of the magnetic energy dissipated through the reconnection. Since the dissipated magnetic energy is finally used to accelerate the outflow, it is considered that
\begin{eqnarray}
\Gamma {t'_{\rm r}\over t'_{\rm dyn}}\propto\rm const.
\end{eqnarray}
This yields
\begin{eqnarray}
\Gamma\propto r^{1/3},
\end{eqnarray}
which is much slower than the fireball acceleration as presented in Eq. (\ref{fbacc}).

\subsubsection{Jet Propagation}
Before the relativistic jet produces the GRB emission, it should first penetrate the thick progenitor medium, which can be a stellar envelope for long GRBs and merger ejecta for short GRBs.
The collision between the jet and the progenitor medium can lead to a forward shock sweeping up the medium and a reverse shock accumulating the jet material. The region between these two shocks is called the jet head, in which the material is very hot and flows quickly and laterally to form a cocoon surrounding the jet. The velocity of the jet head is determined by balancing the ram pressures of the forward-shocked medium and the reverse-shocked jet \citep{2003MNRAS.345..575M}:
\begin{equation}\label{bh}
\beta_{\rm h}=\frac{\beta_{\rm j}\tilde L^{1/2}+\beta_{\rm e}}{\tilde L^{1/2}+1}
\end{equation}
with $\tilde L\simeq{L_{\rm j}}/({\Sigma_{\rm j}\rho_{\rm e}\Gamma_{\rm e}^2c^3})$,
where $\beta_{\rm j}\simeq1$, $L_{\rm j}$, and $\Sigma_{\rm j}$ are the velocity, one-sided luminosity, and cross section of the unshocked jet, respectively, and $\rho_{\rm e}$, $\beta_{\rm e}$, and $\Gamma_{\rm e}$ are the density, velocity, and Lorentz factor of the circum-material.

The high pressure of the cocoon can drive a collimation shock into the jet material toward the jet axis. Therefore, the jet can be gradually collimated and the evolution of the jet cross section is determined by the pressure in the cocoon, which is given by $P_{\rm c}={E_{\rm c}/( 3V_{\rm c}})$. Here the total energy stored in the cocoon $E_{\rm c}$ and the cocoon's volume $V_{\rm c}$ are given by \cite{Bromberg2011}
\begin{eqnarray}\label{Ec}
E_{\rm c}&=&\int_0^{t} L_{\rm j}\left(1-\beta_{\rm h}\right)dt',\\
V_{\rm c}&\sim& \pi r_{\rm c}^2z_{\rm h},
\end{eqnarray}
respectively, where $z_{\rm h}=\int_0^{t}\beta_{\rm h}cdt'$ is the height of the jet head, $r_{\rm c}=\int_0^{t}\beta_{\rm c,\perp}cdt'$, and the lateral expansion velocity reads $\beta_{\rm c,\perp} = [P_{\rm c}/\bar{\rho}_{\rm e}(z_{\rm h})c^2]^{1/2}$ and the average density of the cocoon $\bar{\rho}_{\rm e}=\int\rho_{\rm e}(z)dV/V_{\rm c}$. By defining a critical point $\hat{z}\approx (L_{\rm j}/\pi c P_{\rm c})^{1/2}$ where the collimation shock converges, the jet cross section can be estimated by
\begin{equation}\label{sigma1}
\Sigma_{\rm j}=\left\{
\begin{array}{ll}
\pi\theta_{\rm j0}^2 z_{\rm h}^2,&{~\rm for}~z<\hat{z}/2,\\
\pi\theta_{\rm j0}^2 \left(\hat{z}/ 2\right)^2,&{~\rm for}~z>\hat{z}/2,
\end{array}\right.
\end{equation}
where $\theta_{\rm j0}$ is the initial opening angle of the jet at launching. Approximately, the shape of the jet gradually transforms from conical to cylindrical. The opening angles of the jet head and the cocoon relative to the central engine can be defined as $\theta_{\rm h}=\left({\Sigma_{\rm j}/ \pi z_{\rm h}^2}\right)^{1/2}$ and $\theta_{\rm c}={r_{\rm c}/ z_{\rm h}}$, respectively. As the jet is squeezed by the cocoon, the jet material can be pushed and accelerated.

\subsection{Prompt Emission}
\subsubsection{Internal Dissipation}
The GRB prompt emission can be produced after the jet breaks out from the progenitor medium. One possibility is that the acceleration of the outflow has not finished, and a remarkable fraction of internal energy still exists in the jet after it becomes transparent. In this case, the emission is produced due to the release of the internal energy, which is called photosphere emission, characterized by a quasi black-body spectrum. However, in observation, although such a black-body component has been indeed found in some GRBs, the typical spectrum of GRB prompt emission is non-thermal as described by the Band spectrum. Therefore, it is considered that the prompt emission is more likely to be produced due to the dissipation of the kinetic energy of the jet during the coasting phase. The photosphere emission, which probably appears after the internal dissipation, would be usually weak.

It could be reasonable to consider that a GRB jet is not a continuous fluid, but consists of a series of intermittent shells of very different velocities. Then, the kinetic energy due to the speed difference can be easily dissipated when the shells collide with each other. Let us consider two shells launched by the central engine at an interval of $\delta
t$. The collision between them happens when the later and more rapid one catches up with the previous slower one at the radius of
\begin{eqnarray}
r_{\rm is}
={2\Gamma_{\rm s}^2c\delta
t\over1-(\Gamma_{\rm s}/\Gamma_{\rm r})^2}\approx6\times10^{12}\Gamma_{\rm s,2.5}^2\delta t_{-3}~{\rm
cm},
\end{eqnarray}
where $\Gamma_{\rm s}$ and $\Gamma_{\rm r}$ are the Lorentz factors of the slower and more rapid shells. The above result can be well consistent with the constraint presented in Eq. (\ref{grbradius}), as the light curve variability is determined by the intermittent activity of the central engine. Furthermore, the relatively small value of the internal shock radius favors the fireball model more than the PFD model because the magnetic reconnection could be too slow to complete the outflow acceleration before such a small radius.

Furthermore, by considering of the energy and momentum conservations before and after the collision, the radiation efficiency of the internal shocks (i.e., the conversion efficiency of the kinetic energy to the internal energy) can be constrained to
\begin{eqnarray}
\epsilon
\leq1-2\left(\sqrt{\Gamma_{\rm r}\over\Gamma_{\rm s}}+\sqrt{\Gamma_{\rm s}\over\Gamma_{\rm r}}\right)^{-1}.
\end{eqnarray}
If $\Gamma_{\rm r}$ is not much higher than $\Gamma_{\rm s}$, the radiation efficiency should be much smaller than unity. This indicates the majority of the jet energy should be released in the afterglow phase, which is, however, not always supported by observations. In comparison, the radiation efficiency of a PFD outflow could be much higher than the fireball.

An actual GRB jet could be a hybrid of a fireball and a PFD outflow. In this case, the jet can be accelerated sufficiently quickly and the dissipation radius is determined by internal collisions. Meanwhile, the majority of the jet energy is stored in magnetic fields, the reconnection of which can be triggered by the internal collisions through rapid turbulence. Such a hybrid model was first proposed by Zhang \& Yan (2011), which provides a very competitive explanation for the internal dissipation of GRB jets \citep{Zhang2011}.

\subsubsection{Shocked Material}\label{shockphysics}
As an important mechanism converting kinetic energy into internal energy, a relativistic shock can store up internal energy behind it of a density of \citep{Blandford1976}
\begin{eqnarray}
e'_{\rm sh}=(\Gamma_{\rm sh}-1)(4\Gamma_{\rm sh}-3)n'_{\rm un}m_{\rm p}c^2,
\end{eqnarray}
where $\Gamma_{\rm sh}$ is the Lorentz factor of the shock, which corresponds to the relative velocity between two shells, and $n'_{\rm un}$ is the comoving number density of the unshocked material. Here the shell is assumed to simply consist of protons and electrons, which share the internal energy by equipartition factors of $\varepsilon_{\rm p}$ and $\varepsilon_{\rm e}$, respectively. Furthermore, the numbers of these charge particles would distribute with their energies as a power law: ${dN_{\rm p/e} / d \gamma_{\rm p/e}}\propto \gamma_{\rm p/e}^{-p}$, where $\gamma_{\rm p/e}$ is the Lorentz factor of the random motion of the protons/electrons in the comoving frame of the shocked region, the index $p$ is a free parameter which could be not much higher than 2. While the typical Lorentz factor of the protons is around $\Gamma_{\rm sh}$, the minimum Lorentz factor of the electrons can be determined at \citep{Sari1998}
\begin{eqnarray}
 \gamma_{\rm m}=\varepsilon_{\rm e} \frac{p-2}{p-1} \frac{m_{\rm p}}{m_{\rm e}}(\Gamma_{\rm sh}-1).
\label{eq:8}
\end{eqnarray}

Simultaneously, the shock can also effectively amplify the primordial magnetic field in the shell. By introducing an equipartition parameter of $\varepsilon_{\rm B}$ for the amplified fields, we can write the magnetic field strength as
\begin{eqnarray}
B^{\prime}=\sqrt{ 8 \pi \varepsilon_{\rm B}e'_{\rm sh}}\approx \sqrt{ 32 \pi \varepsilon_{\rm B} \Gamma_{\rm sh}^{2} n'_{\rm un} m_{\rm p} c^{2}}.
\end{eqnarray}
This magnetic field can make the electrons very radiative and change the energy distribution of the electrons. By invoking the total power of the synchrotron radiation of the electrons, a cooling Lorentz factor can be defined as \citep{Sari1998}
\begin{eqnarray}
 \gamma_{\rm c}=\frac{6 \pi m_{\rm e} c}{\sigma_{\rm T} \Gamma_{\rm sh} B^{\prime 2} t},
\label{eq:12}
\end{eqnarray}
where $t$ is the dynamical timescale of the shock. The meaning of $\gamma_{\rm c}$ is that, for $\gamma_{\rm e}>\gamma_{\rm c}$, the electron spectrum has been changed substantially because of the primary energy of these electrons has been radiated. On the contrary, the electron spectrum would keep its original form for $\gamma_{\rm e}<\gamma_{\rm c}$. To be summarized, the final electron spectrum can be written as:\\
(i) fast cooling case ($\gamma_{\rm c}<\gamma_{\rm m}$)
\begin{eqnarray}\label{Edisfast}
{d N_{\rm e}\over d\gamma_{e}}\propto\left\{\begin{array}{lll}
0, & \gamma_{\rm e}<\gamma_{c} \\
\gamma_{\rm e}^{-2}, & \gamma_{\rm c}<\gamma_{\rm e}<\gamma_{\rm m} \\
\gamma_{\rm e}^{-p-1}, & \gamma_{\rm m}<\gamma_{\rm e} ,
\end{array}\right.
\label{eq:elecspec1}
\end{eqnarray}
(i) slow cooling case ($\gamma_{\rm c}>\gamma_{\rm m}$)
\begin{eqnarray}\label{Edisslow}
{d N_{\rm e}\over d\gamma_{e}}\propto\left\{\begin{array}{lll}
 0, & \gamma_{\rm e}< \gamma_{\rm m} \\
 \gamma_{\rm e}^{-p}, & \gamma_{\rm m}<\gamma_{\rm e}<\gamma_{\rm c} \\
 \gamma_{\rm e}^{-p-1}, & \gamma_{\rm c}<\gamma_{\rm e}.
\end{array}\right.
\label{eq:elecspec1}
\end{eqnarray}

\subsubsection{Synchrotron Emission}\label{synchrotron}
The radiation coefficient of the shocked material due to the synchrotron radiation of electrons is given by
\begin{eqnarray}
 j^{\prime}_{\nu^{\prime}}=\frac{1}{4\pi}\int_{\gamma_{\rm m}}{dN_{\rm e}\over d\gamma_{\rm e}} P_{\nu^{\prime}}^{\prime}(\gamma_{\rm e})d\gamma_{\rm e},
 \label{eq:radieffic}
 \end{eqnarray}
where $\nu'$ is the emission frequency measured in the comoving frame and the radiation spectrum of a single electron reads \citep{Rybicki1979}
\begin{eqnarray}
%\begin{aligned}
 P_{\nu^{\prime}}^{\prime}\left(\gamma_{\rm e}\right) = \frac{\sqrt{3} q_{\rm e}^{3} B^{\prime} }{m_{\rm e} c^{2}}\left[\frac{\nu^{\prime}}{\nu_{\rm c}^{\prime}} \int_{\nu^{\prime} /\nu_{\rm c}^{\prime}}^{\infty} K_{5 / 3}(y) d y\right],
 %\\&= \frac{\sqrt{3} q_{\rm e}^{3} B^{\prime} }{m_{\rm e} c^{2}} F\left(\frac{\nu^{\prime}}{\nu_{0}^{\prime}}\right) .
%\end{aligned}
\label{eq:6}
\end{eqnarray}
where $q_{\rm e}$ is the electron's charge and $ \nu'_{\rm c}=3q_{\rm e}B'\gamma_{\rm e}^2/ (4\pi m_{\rm e}c)$.

Because of the quasi-monochromaticity of synchrotron radiation, it can be found that the radiation spectrum of electrons would basically trace the energy distribution of them. Therefore, according to Eqs. (\ref{Edisfast}) and (\ref{Edisslow}), the synchrotron spectrum of the shocked material can be approximated by \citep{Sari1998}: \\
(i) Fast cooling case
 \begin{eqnarray}
 j'_{\nu'}=j_{\nu',\max}\times\left\{ \begin{array}{lll}
 \left({\nu^{'}\over \nu^{'}_{\rm c}}\right)^{1/ 3},& \nu^{'}<\nu^{'}_{\rm c}\\
 \left({\nu^{'}\over \nu^{'}_{\rm c}}\right)^{-1/ 2},& \nu^{'}_{\rm c}<\nu^{'}<\nu^{'}_{\rm m}\\
 \left({\nu^{'}_{\rm m} \over \nu^{'}_{\rm c}}\right)^{-1/ 2} \left({\nu^{'}\over \nu^{'}_{\rm m}}\right)^{-p/ 2},& \nu^{'}_{\rm m}<\nu^{'},
 \end{array}\right.
 \label{eq:16}
 \end{eqnarray}
(i) Slow cooling case
 \begin{eqnarray}
 j_{\nu'}=j_{\nu',\max}\times\left\{ \begin{array}{lll}
 \left({\nu^{'}\over \nu^{'}_{\rm m}}\right)^{1/ 3},& \nu^{'}<\nu^{'}_{\rm m}\\
 \left({\nu^{'}\over \nu^{'}_{\rm m}}\right)^{\left({1-p}\right)/ 2},& \nu^{'}_{\rm m}<\nu^{'}<\nu^{'}_{\rm c}\\
 \left({\nu^{'}_{\rm c} \over \nu^{'}_{\rm m}}\right)^{\left({1-p}\right)/ 2} \left({\nu^{'}\over \nu^{'}_{\rm c}}\right)^{-p/ 2},& \nu^{'}_{\rm c}<\nu^{'},
 \end{array}\right.
 \label{eq:17}
 \end{eqnarray}
where the peak value of the radiation coefficient and the two characteristic frequencies are respectively defined as
\begin{eqnarray}
j'_{\nu',\max}=n'_{\rm e}\frac{m_{\rm e} c^2 \sigma_{\rm T}}{3 q_{\rm e}} B'
\label{eq:18}
\end{eqnarray}
and
\begin{eqnarray}
 {\nu'}_{\rm m/c}=\frac{3 q_{\rm e} B'}{4 \pi m_{\rm e} c} \gamma_{\rm m/c}^{2},
\label{eq:18}
\end{eqnarray}
where the number density of the electrons is given by $n'_{\rm e}=n'_{\rm sh}$.

For an order-of-magnitude analysis of the synchrotron emission of internal shocks, we can write the comoving particle number density of the unshocked shells by
\begin{eqnarray}
n'_{\rm un}={L_{\rm iso}\over 4\pi R_{\rm is}^2\Gamma^2m_{\rm p}c^3},
\end{eqnarray}
where $L_{\rm iso}\sim 10^{50}\rm erg~s^{-1}$ is the isotropic-equivalent luminosity of the jet. Supposing the relative Lorentz factor $\Gamma_{\rm sh}\approx\Gamma_{\rm r}/2\Gamma_{\rm s}$ between different shells is on the order of a few to a few of tens, we can get
\begin{eqnarray}
h{\nu}_{\rm m}=2 C_p^2 \delta t_{-3}^{-1} \Gamma_{2.5}^{-2}\Gamma_{\rm sh,1}^3 L_{\rm iso,50}^{1/2}\epsilon_{\rm B,-2}^{1/2}\epsilon_{e,-1}^2\rm MeV
\end{eqnarray}
which is extremely higher than $h{\nu}_{\rm c}$ (fast cooling case) and well consistent with the peak energy of the prompt GRB emission, where $C_{p}=(p-2)/(p-1)$. However, the problem of this synchrotron emission model is that, the spectrum below the peak ($F_{\nu}\propto\nu^{-1/2}$) is too soft to explain the observed hard spectrum ($F_{\nu}\propto\nu^{0}$). This indicates that more complex factors should be taken into account for reproducing the observed GRB spectrum, e.g., the decay of the magnetic field behind the shock \citep{Peer2006}, the Comptonization of the photosphere emission \citep{Beloborodov2010}, and the temporal evolution of the emission spectrum \citep{Deng2014} etc.

\subsubsection{High-Energy Photon and Neutrino Emission}
Accompanying with the synchrotron radiation, the inverse Compton scattering of the synchrotron photons off relativistic electrons would lead to high-energy emissions above hundreds of GeV. Nevertheless, these high-energy photons are further subjected to severe absorption via two-photon annihilation into electron-positron pairs. The absorption can happen inside the jet or far away from it.
On the one hand, the optical depth of internal attenuation is highly dependent on the bulk Lorentz factor $\Gamma$ of the jet. Only for sufficiently high  $\Gamma$, these high-energy photons can escape from the emitting region, which therefore can be used to constrain the value of $\Gamma$.
On the other hand, those escaping high-energy photons would further interact with cosmic infrared background photons, leading to electron-positron pair production. These secondary pairs can further upscatter the cosmic microwave background, leading to secondary gamma-ray photons, which lag behind the primary high-energy emission.
%These secondary photons reach the observer with a time delay relative to the primary emissions.
Moreover, since the secondary pairs can be deflected by the intergalactic magnetic field, the delayed high-energy emission should also deviate from the direction of the primary emission and thus form a diffuse high-energy halo surrounding the GRB. %Therefore, detecting such delayed secondary photons can provide constraints on the primary spectrum and the intergalactic magnetic field strength.

Because of the intense radiation in the GRB ejecta, the shock-accelerated protons
can lose their energy to produce mesons such as $\pi^0$ and $\pi^\pm$ etc, and subsequently generate neutrinos
by the decay of $\pi^{\pm}$, i.e.,
\begin{eqnarray}
\pi^{\pm}\rightarrow \mu^{\pm}+\nu_{\mu}(\bar{\nu}_{\mu})\rightarrow
e^{\pm}+\nu_{\rm e}(\bar{\nu}_{\rm e})+\bar{\nu}_{\mu}+\nu_{\mu}.
\end{eqnarray}
The timescale of the photomeson processes can be calculated by \cite{1997PhRvL..78.2292W}
\begin{eqnarray}
{t'}_{\pi}^{-1}   =  {c\over 2{\gamma}_{\rm p}^2}\int_{\tilde{E}_{\rm
th}}^{\infty}\sigma_{\pi}(\tilde{E})\xi(\tilde{E})\tilde{E}
 \left[\int_{\tilde{E}/2\gamma_{\rm p}}^{\infty}n'(E'_{\gamma})
{E'}_{\gamma}^{-2}dE'_{\gamma}\right]d\tilde{E},\label{tpg1}
\end{eqnarray}
where $\gamma_{\rm p}$ is the proton's Lorentz factor, $\sigma_{\pi}(\tilde{E})$ is the cross section of photopion
interactions for a target photon with energy $\tilde{E}$ in the
proton's rest frame, $\xi$ is the inelasticity defined as the
fraction of energy loss of a proton to the resultant pions, and
$\tilde{E}_{\rm th}=0.15\rm GeV$ is the threshold energy of the
interactions. Then, the fraction of the energy loss of the protons to pions can be written as
\begin{equation}
f_{\pi}\approx 1-\exp\left(-{t'\over {t'}_{\pi}}\right)\approx \min\left[\Gamma t/{t'}_{\pi},1\right],\label{fpiapp}
\end{equation}
where $t$ is the observer's time. By taking a constant ratio between different pions as $\pi^{\pm}:\pi^0=2:1$, the time-integrated muon-neutrino spectrum can be given by
\begin{equation}
E_{\nu}^2\phi_{\nu}\equiv  {1\over 4\pi
d_{\rm L}^2}E_{\nu}^2{dN_{\nu}\over dE_{\nu}}={1\over 4\pi
d_{\rm L}^2}{f_{\pi}\over 3} {E}_{\rm p}^2{dN_{\rm p}\over
dE_{\rm p}}\label{nuspectra},
\end{equation}
where $d_{\rm L}$ is the luminosity distance of the burst and the neutrino energy is related to the primary proton's energy by $E_{\nu}={1\over4}\xi E_{\rm p}$ since the two
resultant muon-neutrinos from the decay of a $\pi^{\pm}$ could
inherit half of the pion's energy roughly evenly. Here, the energy distribution of the shock-accelerated protons is expected to have a form of $({dN_{\rm p}/ dE_{\rm p}})\propto{E}_{\rm p}^{-2}$, where the proportional
coefficient can be calculated by $\epsilon_{\rm p}E/\ln(E_{\rm p,\max}/E_{\rm p,\min})$. In addition, due to the presence of the stochastic magnetic fields, the
ultra-high energy pions and muons can also lose their energy via
synchrotron radiation before decay, which therefore steepens the neutrino spectrum at very high energies by timing an extra power law as $E_{\nu}^{-2}$.

\subsection{Multi-Wavelength Afterglows}
\subsubsection{External Reverse Shock}
\begin{figure}
\begin{center}
	\includegraphics[width=\textwidth]{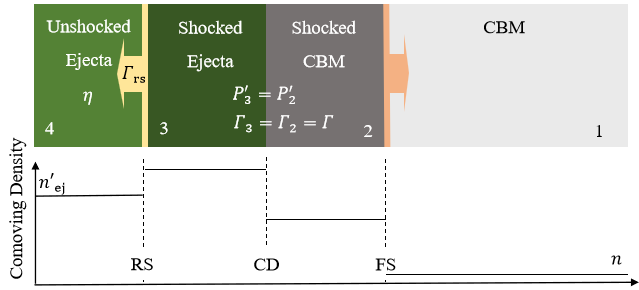}
	\caption{Illustration of the external reverse and forward shocks.}\label{FRS}
	\end{center}
\end{figure}

After the internal shock processes, all the jet material would eventually merge into a whole ejecta of a Lorentz factor $\eta$, which moves into the circum-burst medium (CBM) persistently. Typically, the CSM can be the interstellar medium (ISM) or a wind environment produced by the progenitor star. Then, the interaction between the ejecta and the CBM can drive a forward shock sweeping up the
CBM and a reverse shock crossing the ejecta, as shown in Fig. \ref{FRS}. This  situation is like the jet propagation in the progenitor material, but the difference is that the shocks here are radiative rather than adiabatic, which therefore can produce emission directly.

By considering that the GRB ejecta has a very small initial thickness of $\Delta_0\sim 10^6$ cm (before spreading), the external reverse shock can only last a limited period of \citep{Sari1995}
\begin{eqnarray}
t_{\rm rs}\sim{\Delta_0\over c}\eta f^{1/2},
\end{eqnarray}
where
\begin{eqnarray}\label{RFequ}
f\equiv{n'_{\rm ej}\over
n}={(4\Gamma+3)(\Gamma-1)\over(4\Gamma_{\rm rs}+3)(\Gamma_{\rm rs}-1)},
\end{eqnarray}
which represents the mechanical equilibrium between the two shocks, where $n'_{\rm ej}$ is the comoving density of the unshocked ejecta, $n$ is the density of the CBM, $\Gamma$ is the Lorentz factor of the shocked region, and
\begin{eqnarray}\label{gamrs}
\Gamma_{\rm rs}={1\over2}\left({\Gamma\over \eta}+{\eta\over\Gamma}\right)
\end{eqnarray}
is the Lorentz factor of the shocked region measured in the comoving frame of the ejecta.
Solving Eqs. (\ref{RFequ}) and (\ref{gamrs}), the Lorentz factor of the shocked region can be derived to \citep{Sari1995}
\begin{eqnarray}
\Gamma=\left\{
\begin{array}{ll}
{1\over\sqrt{2}}\eta^{1/2}f^{1/4},&{\rm for~}f\ll\eta,~~({\rm relativistic~case})\\
\eta(1-\sqrt{2\xi}),&{\rm for~}f\gg\eta,~~({\rm newtonian~case})
\end{array}\right.\label{c1_mod_rfgamma}
\end{eqnarray}
where $\xi=4\eta^2/(7f)$. For typical parameter values, we can usually obtain $\Gamma_{\rm rs}\ll\Gamma$, which leads to the reverse shock emission peaks in the optical bands while the forward shock is in X-rays. The duration of the optical flash due to the reverse shock is about $t_{\rm rs}\sim 100$ s, after which the flux decays as $t^{-2}$.

The superposition of the reverse and forward shock emissions would lead the early optical afterglow to be more complicated than that in the other bands. When the reverse shock emission is strong and dominated at the very early stage, a rapid rising and decaying optical flash would show up, which is gradually followed by a normal decay at a late time or by a rebrightening signature due to the forward shock emission.
\subsubsection{External Forward Shock}
After the external reverse shock crosses the GRB ejecta, the external forward shock still exists and moves persistently, which can enter into a self-similar evolution \citep{Blandford1976}. Then, by taking a thin shell approximation, the dynamical evolution of the forward shock can be easily determined by the energy conservation law as
\begin{eqnarray}\label{ESeconsv}
dE_{\rm k}=-\varepsilon\Gamma(\Gamma-1)c^2 dM_{\rm sw},
\end{eqnarray}
where $\varepsilon$ represents the radiation efficiency of the shock, $M_{\rm sw}$ is the mass of the swept-up CBM, and the kinetic energy of the system can be expressed as
\begin{eqnarray}\label{}
E_{\rm k}=(\Gamma-1) (M_{\rm ej}+M_{\rm sw})c^2+(1-\varepsilon)\Gamma(\Gamma-1)M_{\rm sw}c^2,
\end{eqnarray}
where $M_{\rm ej}$ is the mass of the ejecta. From Eq. (\ref{ESeconsv}), a differential dynamical equation of the shock can be obtained as \citep{{Huang1999,Huang2000}}
\begin{eqnarray}\label{ESdyn}
 \frac{d \Gamma}{d M_{\mathrm{sw}}}=-\frac{\Gamma^{2}-1}{M_{\mathrm{ej}}+\varepsilon M_{\mathrm{sw}}+2(1-\varepsilon) \Gamma M_{\mathrm{sw}}}.
\end{eqnarray}
The deceleration of the forward shock can be significant only when the mass of the swept-up CBM is much larger than $M_{\rm ej}/\eta$.

For an adiabatic approximation, we can solve from Eq. (\ref{ESdyn}) to $\Gamma\propto M_{\rm sw}^{-1/2}$ for $M_{\rm ej}/\eta \ll M_{\rm sw}\ll \eta M_{\rm ej} $ and $\beta\propto M_{\rm sw}^{-1/2}$
for $M_{\rm sw}\gg \eta M_{\rm ej} $, where $\beta=\sqrt{1-\Gamma^{-2}}$. Here the evolution of the mass of the swept-up CBM is determined by
\begin{eqnarray}\label{Msw}
\frac{d M_{\rm{sw}}}{d r}=4\pi r^{2} n m_{p},
\end{eqnarray}
and
\begin{eqnarray}\label{rt}
 {dr\over dt}={\beta c\over 1-\beta},
\label{eq:31}
\end{eqnarray}
where $r$ is the radius of the shock and $t$ is the time in the observer's frame. The term $(1-\beta)$ appearing in the above equation is due to the Doppler effect compressing the local dynamical time of the shock for the observer. Then, for an ultra-relativistic case, we can approximately take $r\sim 2\Gamma^2ct$.

\begin{figure}
\begin{center}
	\includegraphics[width=0.7\textwidth]{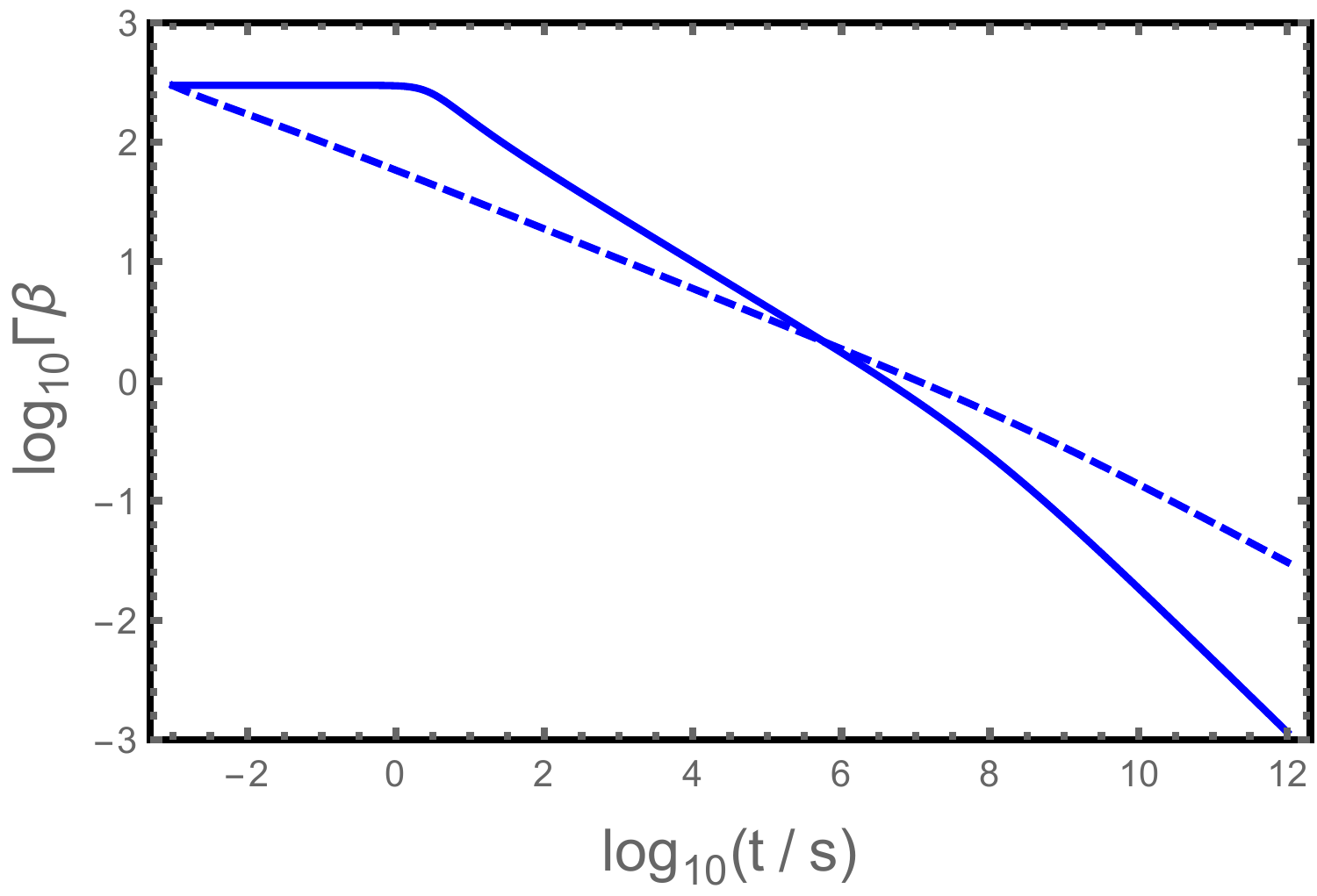}
	\caption{Dynamical evolution of a GRB forward shock in the ISM (solid) and wind (dashed) environments for $\varepsilon=0$.}\label{ESdyn}
	\end{center}
\end{figure}

By further taking the CBM density as $n=Ar^{-k}$ ($k=0$ for ISM and $k=2$ for wind environment), the dynamical evolution of the external forward shock can be solved to
\begin{eqnarray}
\Gamma=\eta \left({t\over t_{\rm dec}}\right)^{-(3-k)/[2(4-k)]}
\end{eqnarray}
for the relativistic case and
\begin{eqnarray}
\beta=\left({t\over t_{\rm nr}}\right)^{-(3-k)/(5-k)}
\end{eqnarray}
for the non-relativistic case. The deceleration timescale $t_{\rm dec}$ and the transition time $t_{\rm nr}$ from the relativistic to the non-relativistic phase are determined by the conditions of $M_{\rm sw}=M_{\rm ej}/\eta$ and $M_{\rm sw}=\eta M_{\rm ej} $, respectively, which yield
\begin{eqnarray}
t_{\rm dec}&=&\left[{(3-k)E_{\rm iso}\over 2^{5-k}\pi A m_{\rm p}\eta^{2(4-k)}c^{5-k}}\right]^{1/(3-k)}\nonumber \\
&=&\left\{
\begin{array}{ll}
5E_{\rm iso,52}^{1/3}\eta_{2.5}^{-8/3}n_0^{-1/3}{\rm ~s},~~{\rm for}~~ k=0,\\
0.003E_{\rm iso,52}\eta_{2.5}^{-4}A_{35.5}^{-1}{\rm ~s},~~{\rm for}~~ k=2,
\end{array}\right.
\end{eqnarray}
and
\begin{eqnarray}
t_{\rm nr}&=&\left[{(3-k)E_{\rm iso}\over 4\pi A m_{\rm p}c^{5-k}}\right]^{1/(3-k)}\nonumber \\
&=&\left\{
\begin{array}{ll}
4\times10^7E_{\rm iso,52}^{1/3}n_0^{-1/3}{\rm ~s},~~{\rm for}~~ k=0,\\
6\times10^7E_{\rm iso,52}A_{35.5}^{-1}{\rm ~s},~~{\rm for}~~ k=2,
\end{array}\right.
\end{eqnarray}
where $M_{\rm ej}=E_{\rm iso}/\eta c^2$ is used. These characteristic timescales can be found easily from the numerical calculation results presented Fig. \ref{ESdyn}. Based on these dynamical results, the synchrotron emission of the forward shock can be calculated by using the formulae presented in previous sections.

\begin{figure}
\begin{center}
	\includegraphics[width=0.9\textwidth]{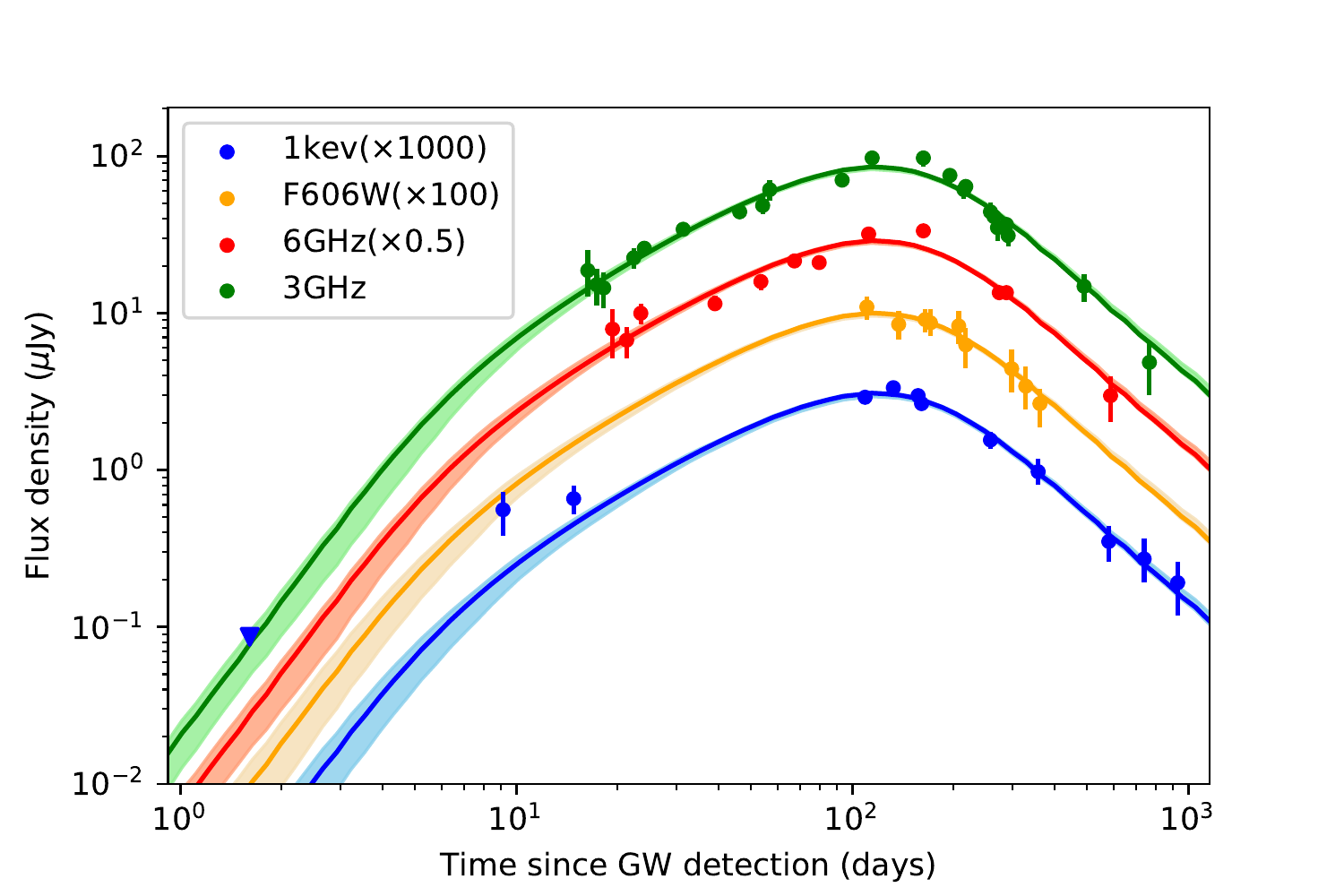}
	\caption{Fitting to the multi-wavelength afterglow light curves of GRB 170817A with an off-axis Gaussian-structured jet.}\label{GRB170817A}
	\end{center}
\end{figure}

The above calculations assume the GRB jet has a top-hat structure, and thus it can be treated isotropically. However, an actual jet is very likely to have an angular structure, which could significantly influence the afterglow observation when the GRB is observed off-axis. In this case, the dynamical evolution for different directions should be calculated separately. The angle-dependence of the Doppler effect also needs to be considered as
\begin{eqnarray}
 {dr_{\theta}\over dt}={\beta_\theta c\over 1-\beta_\theta\cos\alpha},
\label{eq:31}
\end{eqnarray}
where the subscript $\theta$ represents the emission element deviates from the symmetric axis by an angle of $\theta$, and the viewing angle of
this element is given by
\begin{eqnarray}
\cos\alpha =\cos\theta\cos\theta_{\rm obs}+\sin\theta\sin\theta_{\rm obs}\cos\varphi,
\label{eq:30}
\end{eqnarray}
where $\theta_{\rm obs}$ is the angle between the jet axis and the line of sight. The total flux of the afterglow emission of a structured jet can be obtained by
\begin{eqnarray}
 F_{\rm \nu}(t)={1\over d_{\rm L}^{2}}\int_{0}^{\pi/2}\int_{0}^{2\pi} \frac{I^{'}_{\nu^{'}}(r,\theta,\varphi)}{\Gamma_{\theta}^3(1-\beta_{\theta}\cos\alpha)^3}d\varphi d\theta,
\label{eq:Fnu}
\end{eqnarray}
where $d_{\rm L}$ is the luminosity function of the GRB and the radiation intensity contributed by the element in the direction $ (\theta,\phi)$ can be written as
\begin{eqnarray}
I'_{\nu'}(r,\theta,\varphi)d\varphi d\theta={j'_{\nu'}(r,\theta,\varphi)\over 4\pi}{M_{\rm sw,\theta}\over 4\pi n'_{\rm sh}m_{\rm p}}d\varphi d\theta,
\end{eqnarray}
where $M_{\rm sw,\theta}$ is an isotropically-equivalent value. Finally, the calculating result of Eq. (\ref{eq:Fnu}) can be used to fit the observational afterglow data, as presented in Fig. \ref{GRB170817A}.

\subsubsection{Post-Standard Afterglow Models}
The observed GRB afterglows exhibit a large variety and complexity in their light curves, although the standard external shock model can explain the general behavior of the light curves. This indicates that some complex factors need to be invoked in the model, e.g., a complicated angular and radial structure of the GRB jet and a possible evolution of the microphysical parameters. Nevertheless, the most critical factor affecting the afterglow emission comes from the central engine, which could still be very active after the prompt phase. Such a hypothesis is strongly supported by the observed X-ray flares and plateaus observed in the afterglow phase. Specifically, a long-lasting active engine could be an accreting BH or a spinning-down NS.

In the BH case, the rate of the fallback accretion starting from the time of $\sim0.1\,\rm s$ would behave as a power law $\dot{M}_{\rm fb}\propto\,t^{-5/3}$. By assuming a feedback efficiency $\eta_{\rm fb}$, the outflow luminosity can be written as
\begin{eqnarray}
L_{\rm fb}&=&\eta_{\rm fb}\dot{M}_{\rm fb}c^{2}\\
&\approx&2\times10^{51}{\rm erg\,s^{-1}}\left({\eta_{\rm
fb}\over0.1}\right)\left({\dot{M}_{\rm
fb,i}\over10^{-3}M_{\odot}\,\rm s^{-1}}\right)\left(t\over
0.1\rm\,s\right)^{-5/3},
\end{eqnarray}
where the subscript `i' stands for ``initial''. For this function, however, the primary energy is released at a very early time so that the later afterglow emission cannot be influenced substantially. On the contrary, in the spinning-down NS case, the energy release can nearly keep constant for a period of
\begin{equation}\label{eq8}
t_{\rm sd}=\frac{3Ic^3}{B_{\rm{p}}^{2}R_{\rm{s}}^{6}}\left (\frac{2\pi}{P_{\rm{i}}}\right)^{-2}=2\times 10^3I_{45}R_{\rm{s,}6}^{-6}B_{\rm{p,}15}^{-2}P_{\rm i,-3}^{2}\,\rm{s},
\end{equation}
where $B_{\rm{p}}$, $R_{\rm{s}}$, and $P_{}$ are the polar magnetic field strength, radius, and spin period of the NS, respectively. Here the NS is assumed to be a millisecond magnetar. The corresponding luminosity is determined by the magnetic dipole radiation as
\begin{equation}\label{eq6}
L_{\rm{md,i}}=\frac{B_{\rm{p}}^{2}R_{\rm{s}}^{6}}{6c^3}\left ( \frac{2\pi}{P_{\rm i}} \right) ^4=9.6\times 10^{42}B_{\rm{p,}12}^{2}R_{\rm{s,}6}^{6}P_{\rm i,-3}^{-4}\rm{erg\ s}^{-1}.
\end{equation}
The complete behavior of the spin-down luminosity can be written as
\begin{equation}\label{sdlum}
	L_{\rm{md}} (t)=L_{\rm{md,i}}\left (1+\frac{t}{t_{\rm{sd}}}\right)^{-2}.
\end{equation}
Sometimes, if the magnetic field of the NS is much lower than $10^{15}$ G and the ellipticity of the NS is high enough, the gravitational radiation of the NS would also play a role in braking the stellar rotation, which determines a spin-down timescale of
\begin{equation}\label{eq7}
t_{\rm{sd}}=\frac{5P_{\rm{i}}^{4}c^5}{2048\pi ^4GI\epsilon ^2 }=9.1\times 10^5\epsilon _{-4}^{-2}I_{45}^{-1}P_{\rm i,-3}^{4}\,\rm{\rm s},%\ \rm{s}\\&\rm{for}\ \alpha =1,
\end{equation}
where $\epsilon$ is the ellipticity. In any case, the constant energy release of the NS before the spin-down timescale can substantivally slow the deceleration of the external shock of the GRB ejecta, which provides a natural explanation of the shallow-decaying light curves of some GRB afterglows \cite{Dai1998a,Dai1998b,Zhang2001}. The dynamical evolution of the external shock with an energy injection can be obtained by substituting Eq. (\ref{sdlum}) into (\ref{ESeconsv}).

In more detail, the energy released from the central engine could be in the form of a Poynting flux initially and finally evolve into a relativistic wind consisting of electrons and positrons. When such a relativistic wind collides with the swept-up medium, a termination shock can be formed and propagate into the wind. In the NS case, the shocked wind region can contribute an extra emission component for the GRB afterglow, just like the emission of a pulsar wind nebulae existing in some supernova remnant \cite{Dai2004,Yu2007}. The existence of such an internal-origin afterglow is helpful for understanding the complexity of GRB afterglow light curves \cite{Yu2010} including some unusual plateaus followed by an extremely sharp decay.

\subsection{Supernova and Kilonova}
\subsubsection{Supernova}
Since long GRBs originate from the collapse of massive stars, they are expected to be associated by supernova emission, which reaches a peak at a time of
\begin{eqnarray}
{t}_{\rm d}&=&\left(\frac{3\kappa{M}_{\rm ej}}{4\pi{cv}}\right)^{1/2}\nonumber\\&=&46\left({\kappa\over 0.2\rm cm^2g^{-1}}\right)^{1/2}\left({{M}_{\rm ej}\over 5M_{\odot}}\right)^{1/2}v_9^{-1/2}\rm day.
\end{eqnarray}
This peak time is determined by the diffusion timescale of photons in the supernova ejecta, where $\kappa$, $M_{\rm ej}$, and $v$ are the opacity, mass, and expanding speed of the supernova ejecta. As usual, the supernova emission is powered by the radiative decays of $^{56}$Ni to $^{56}$Co and to $^{56}$Fe. The rate of the energy release of these processes is given by
\begin{equation}\label{eqPowrad}
	\dot{q}_{\rm r}=\left(\epsilon_{\rm Ni}-\epsilon_{\rm Co}\right) {\rm e}^{-t/\tau_{\rm Ni}}+\epsilon_{\rm Co} {\rm e}^{-t/\tau_{\rm Co}},
\end{equation}
where $\epsilon_{\rm Ni}=3.90 \times 10^{10} \rm {erg} ~{s}^{-1} {~g}^{-1}$, $\epsilon_{\rm Co}=6.78 \times 10^{9}\rm {erg}~ {s}^{-1} {~g}^{-1}$, $\tau_{\rm Ni}=8.76 \rm {day}$, and $ \tau_{\rm Co}=111.42\rm {day}$.
The light curve of the supernova emission is determined by the radiative transfer of the thermal energy in the supernova ejecta:
\begin{equation}
	{L}=-4 \pi {r}^{2} \frac{{c}}{3 \kappa \rho} \frac{\partial {u}}{\partial {r}}.
	\end{equation}
where $r$ is the radius, $u$ and $\rho$ are the energy and mass densities.

For an order-of-magnitude analysis, by taking $\frac{\partial {u}}{\partial {r}} \sim \frac{{u}}{{R}} \sim \frac{U}{{VR}}$,
the bolometric luminosity of the supernova can be estimated as follows
\begin{equation}\label{eqSNLe}
	{L}_{\rm e}=\frac{{Uc}}{{R} {\tau}}\left(1-{e}^{-\tau}\right),
\end{equation}
where $U$ is the total internal energy, $V$ is the volume, $R$ is the surface radius, and $\tau=\kappa \rho R$ is the optical depth of the ejecta.	The evolution of the internal energy can be determined by the energy conservation of the ejecta as \citep{Arnett1982,Kasen2010}
\begin{equation}\label{eqSNenergy}
	\frac{{dU}}{{dt}}=M_{\rm Ni}\dot{q}_{\rm r}-{L}_{\rm e}-{P} \frac{\partial {V}}{\partial {t}},
\end{equation}
where $M_{\rm Ni}$ is the total mass of $^{56}$Ni and $P=U/3V$ is the pressure. The term $-PdV$ represents the adiabatic cooling of the ejecta. The combination of Eq. (\ref{eqSNLe}) and (\ref{eqSNenergy}) can yield a bolometric light curve for the supernova\footnote{The light curve can also derive from the following integral \cite{Arnett1982}:
\begin{equation}
L_{\rm e}(t)= 2M_{\rm Ni}\left(\int_{0}^{\tilde{t}} \dot{q}_{\rm r}\tilde{t}'e^{\tilde{t}'^{2}} d\tilde{t}'\right)e^{-\tilde{t}^{2}},
\end{equation}
where $\tilde{t}=t/t_{\rm d}$.}. Sometimes, an extra power could be involved in Eq. (\ref{eqSNenergy}), e.g., if the supernova ejecta can also absorbed energy from the central engine. Finally, by assuming a black-body spectrum for the supernova emission, an effective temperature of the supernova emission can be defined as $T_{\rm BB}=({L_{\rm e}/ 4\pi R_{\rm ph}^2\sigma})^{1/4}$, where $R_{\rm ph}$ is the radius of the photosphere, which corresponds to $\kappa\rho(R-R_{\rm ph})=1$. Then, the chromatic luminosity for a specific frequency $\nu$ can be given by
\begin{equation}
	L_{\nu}={8\pi^2R_{\rm ph}^2\over c^2}{h\nu^3\over {\rm e}^{h\nu/kT_{\rm BB}}-1},
\end{equation}
which can be used to compare with the observational magnitudes of the supernova.

\subsubsection{Kilonova/Mergernova}
During a merger of double NSs or an NS and a BH, a non-relativistic mass of $\sim10^{-4}-10^{-2}M_\odot$ can be ejected more widely than the GRB jet due to the effects of tidal disruption, collision squeeze, and accretion feedback. It is suggested that nearly half of the elements heavier than iron in the universe can be synthesized in this neutron-rich merger ejecta, through the rapid neutron-capture process (r-process) \citep{Lattimer1974,Lattimer1976}.
Then, like the supernova situation, the radioactive decays of the $r$-process elements can also lead to a transient thermal emission (i.e., kilonova), by heating the merger ejecta with a power of \citep{Lippuner2015}
\begin{equation}\label{eq9}
\dot{q}_{\mathrm{r}}=4 \times 10^{18}\left[\frac{1}{2}-\frac{1}{\pi} \arctan \left (\frac{t-t_{0}}{\sigma}\right)\right]^{1.3} \rm erg~ s^{-1} g^{-1}
\end{equation}
and a thermalization efficiency of \citep{Barnes2016}
\begin{equation}\label{eq10}
	\eta_{\mathrm{th}}=0.36\left[\exp \left (-0.56 t_{\mathrm{day}}\right)+\frac{\ln \left (1+0.34 t_{\mathrm{day}}^{0.74}\right)}{0.34 t_{\mathrm{day}}^{0.74}}\right]
\end{equation}
where $ t_0=1.3\,{\rm s} $, $ \sigma=0.11\,{\rm s} $, and $t_{\rm day} = t/{\rm day}$.

The kilonova emission calculation is in principle the same as the case of supernova emission, but with different power and different ejecta. The peak emission of kilonovae is expected to appear at
\begin{eqnarray}
{t}_{\rm d}=5\kappa_{1}^{1/2}M_{\rm ej,-2}^{1/2}v_{10}^{-1/2}\rm day.
\end{eqnarray}
Here a high reference value of $\sim 10\rm cm^2g^{-1}$ is taken for the opacity, by considering that the opacity can be significantly increased due to the formation of a large number of lanthanide elements \cite{Barnes2013}. For a reference luminosity $L_{\rm kn}\sim 10^{41}\rm erg~s^{-1}$, we can determine the peak wavelength of the kilonova emission to
\begin{eqnarray}
\lambda_{\rm p}={hc\over 5k_{\rm B}}\left[{L_{\rm kn}\over 4\pi (v t_{\rm d})^2\sigma}\right]^{-1/4}=1.6\kappa_{1}^{1/4}L_{\rm kn,41}^{-1/4}M_{\rm ej,-2}^{1/4}v_{10}^{1/4}\rm \mu m,
\end{eqnarray}
which indicates the kilonova emission is inclined to be red. Nevertheless, the different components of the merger ejecta could have very different electron fractions, which can substantially affect the efficiency of the r-processes and as well as the opacity of the ejecta. Therefore, besides the red emission component, the kilonova emission can, in principle, also include relatively blue components. Furthermore, the spectrum and light curve of kilonovae can also be sensitive to the viewing angle because of the high anisotropic structure of the merger ejecta (especially for the NS and BH mergers) \citep{Zhu2020}.

\begin{figure}
\begin{center}
 	\includegraphics[width=0.9\textwidth]{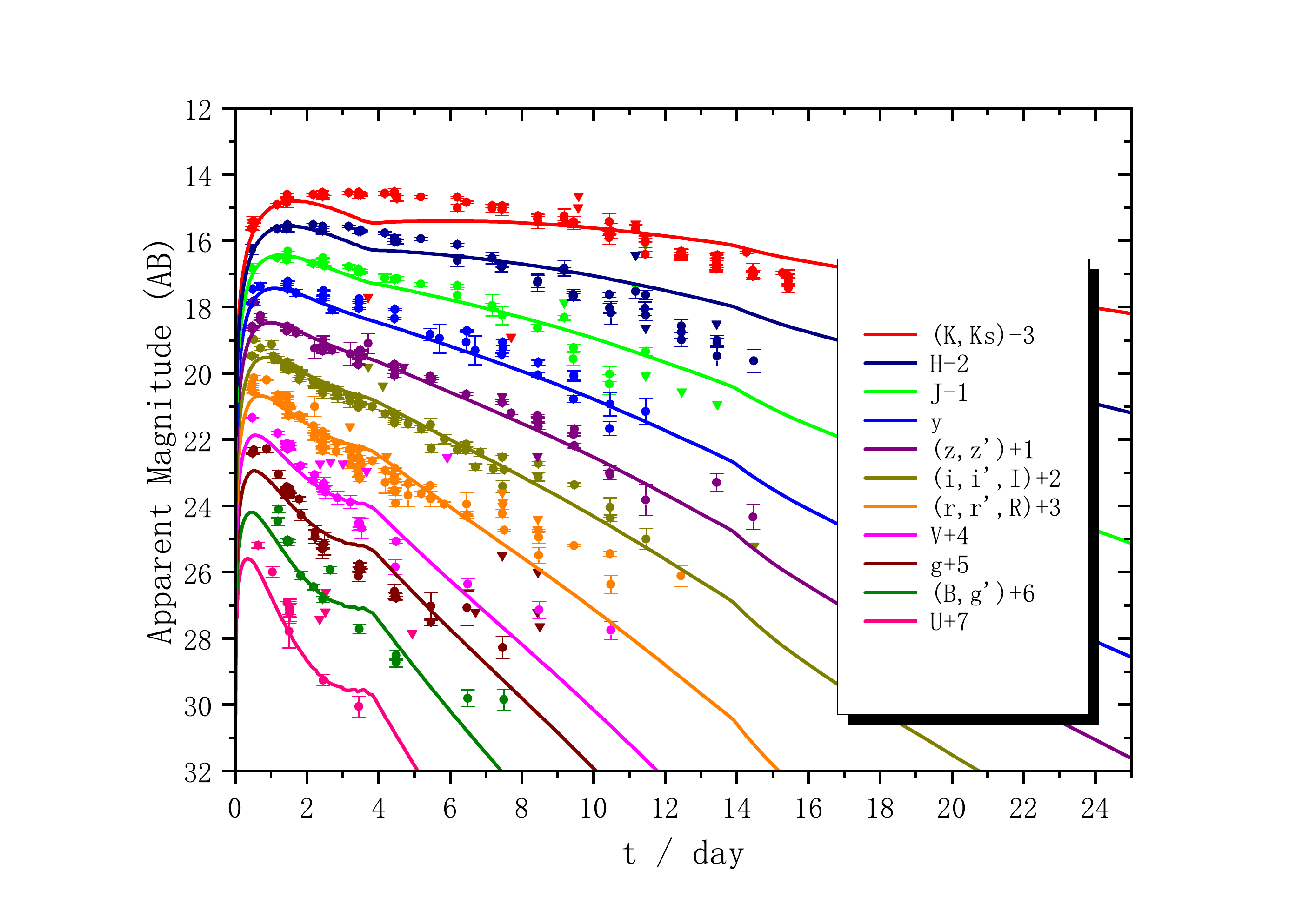}
	\caption{The fitting to the light curves of kilonova AT2017gfo with hybrid energy sources inlcuding a pinning-down NS and the radioactivity of r-process elements (solid lines). From \cite{2018ApJ...861..114Y}}\label{AT2017gfo}
\end{center}
\end{figure}

Finally, for double NS mergers, one of the most concerned topics is the nature of the merger product. As implied by the afterglows of short GRBs, the merger product could be a long-lived massive NS \citep{Dai2006,Giacomazzo2013}. In this case, the kilonova emission can be extra powered by the relativistic wind from the NS, and thus its luminosity could be significantly enhanced\cite{Yu2013,Metzger2014}. Therefore, it was suggested that the transient thermal emission during a merger event could be generally called a mergernova \citep{Yu2013}, instead of a kilonova. For example, the relatively high luminosity of kilonova AT 2017gfo is probably powered by hybrid energy sources \cite{2018ApJ...861..114Y}, since the pure radioactive scenario requires too much high ejecta mass to be accounted for by a double NS merger. Fitting to the observational data, as presented in Fig. \ref{AT2017gfo}, give a stringent constraint on the property of the remnant NS, which indicates the complexity of the early evolution of such newborn NSs. In addition, the interaction between the NS wind and merger ejecta can further contribute to a non-thermal emission component, which is detectable when the merger ejecta becomes transparent \citep{Gao2016,Wu2021}.

%---------------------------------------------------------------------------------------------------------------%
%---------------------------------------------------------------------------------------------------------------%
%---------------------------------------------------------------------------------------------------------------%
%---------------------------------------------------------------------------------------------------------------%

\section{Statistics and Cosmological Applications}
\subsection{Luminosity Function}
In a statistical view, by denoting the event rate of GRBs and their luminosity function as $\dot{R}_{\rm GRB}$ and $\Phi(L)$, the detectable numbers of GRBs in different flux and redshift ranges can be calculated by \cite{2020ApJ...902...83T}
\begin{eqnarray}
N(P_1, P_2)&=&{\Delta\Omega\over 4\pi} T \int_{0}\int^{P_{2}}_{P_{1}}\int_0\eta(P)\nonumber\\
&\times&\dot{R}_{\rm GRB }(z) \Phi_{\rm}(L)\sin\theta_{\rm v}~d \theta_{\rm v} dP{dV(z)\over 1+z},
\label{EQN: PPFD}
\end{eqnarray}
and
\begin{eqnarray}
N(z_1, z_2)&=&{\Delta\Omega\over 4\pi} T \int^{z_2}_{z_1}\int_{0}\int_0\eta(P)\vartheta_z(z,P) \nonumber\\
& \times&\dot{R}_{\rm GRB }(z)\Phi_{\rm }(L)\sin\theta_{\rm v}~d\theta_{\rm v}dP{dV(z)\over1+z}, \label{EQN: RD}
\end{eqnarray}
where $\Delta \Omega$ is the field of view of a telescope, $T$ is the working time with a duty cycle of $\sim$50\%, $\eta(P)$ and $\vartheta(z,P)$ are the trigger efficiency and the probability of redshift measurement, respectively, and $dV(z)$ is the comoving cosmological volume element. Here the integral over the viewing angle $\theta_{\rm v}$ is included, because the GRB luminosity could be direction-dependent if the jet has a significant angular structure. By connecting the GRB rates to the cosmic star formation rates (SFRs), we can use the above equations to model the observational distributions and constrain the luminosity function. As usual, without considering the jet structure, the luminosity function is found to have a broken-power-law form. However, when a Gaussian jet structure is invoked, it can be found that a single-power-law luminosity function would become a better choice \cite{2020ApJ...902...83T}. This hints that the usual low-energy power law in the luminosity function could be a result of off-axis observations.

\subsection{High-redshift Universe}
\begin{figure}
\begin{center}
	\includegraphics[width=0.9\textwidth]{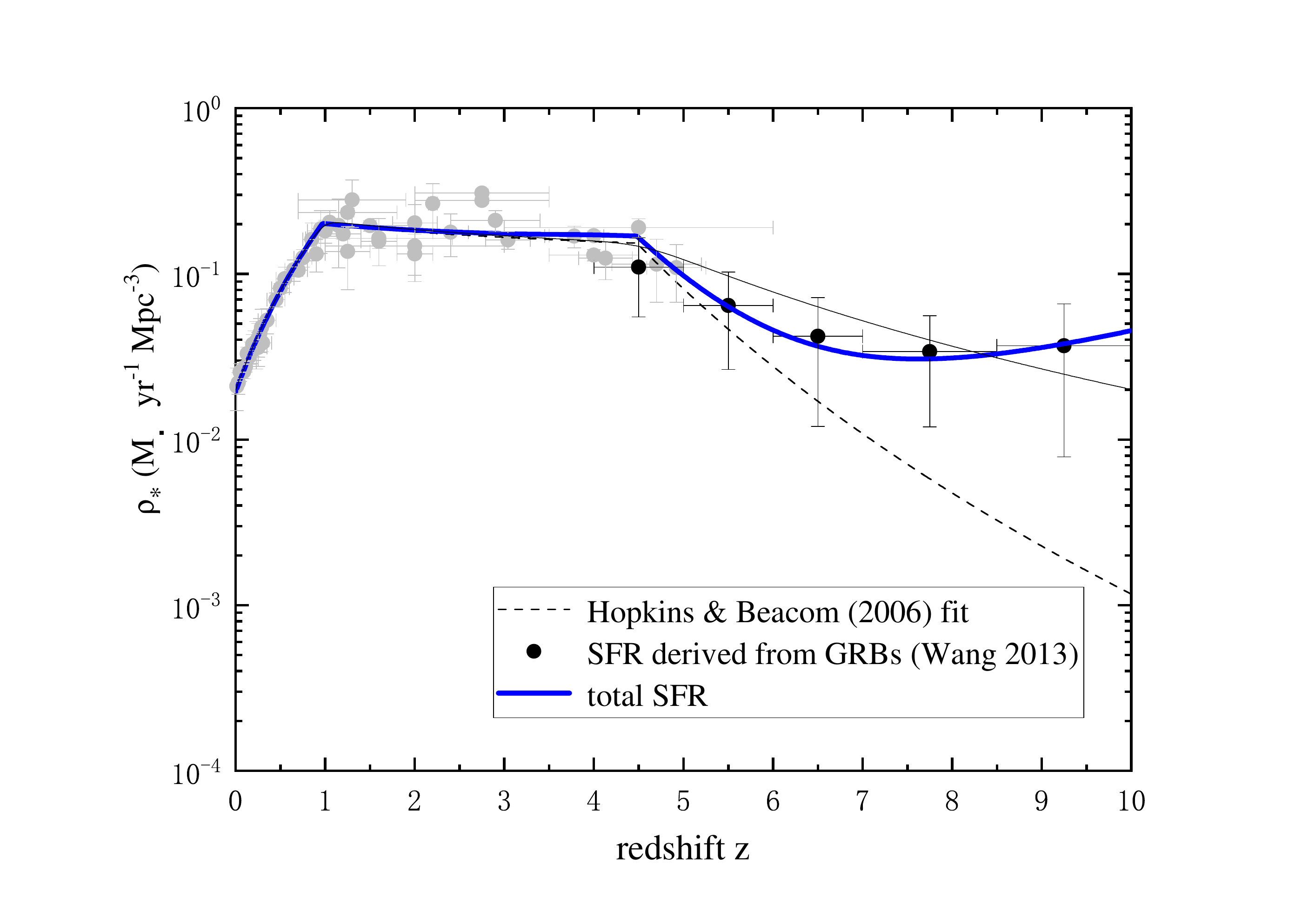}
	\caption{Cosmic star formation history inferred from high-redshift GRBs. From \cite{Wang2013}}\label{SFR}
	\end{center}
\end{figure}

The association of long GRBs with core-collapse supernovae indicates that the GRB event rates could trace the cosmic star formation history either unbiasedly or, more probably, with an additional evolution effect. Therefore, they can provide a complementary technique to measure the star formation rate at high redshifts where direct measurement is extremely difficult \citep{2009ApJ...705L.104K,Wang2009}, as presented in Fig. \ref{SFR}. Here the selection effect and calibration from the GRB rate to SFR must be properly handled.

Some works find that the rate of long GRBs has an excess compared to SFR at high redshifts \cite{2009ApJ...705L.104K,Wang2013}. Possible reasons includes metallicity evolution \cite{Li2008}, the evolution of the stellar initial mass function \cite{Wang2011}, and evolving luminosity function break \cite{Virgili2011}. Theory and observation both support that long GRBs prefer to explode in a low-metallicity site, leading to more GRBs at a high-redshift universe with low metals. For evolving initial mass function, ``top-heavy" initial mass function would lead to more massive
stars in the early universe which can form much more GRBs.

Absorption processes imprinted on the spectra
of GRBs are the main sources of information about the
chemical properties of the high-redshift universe.
The progenitors of long GRBs are believed to be massive stars, so their
number is abundant in the early universe. The damped
Lyman-$\alpha$ and absorption lines in GRB spectra can be used to probe the metal
enrichment history \cite{Wang2012}, and cosmic reionization \cite{Totani2006}. The forthcoming James Webb Space Telescope
would detect afterglow spectra of GRBs out to $z>10$, shielding light on the process of
cosmic reionization.

\subsection{Luminosity Correlations of GRBs}
Since GRBs can be observed much more distant than SNe Ia, they can fill the gap between SNe Ia and cosmic microwave background (CMB) in cosmological studies. Similar to SNe Ia, it has been proposed to use GRB correlations to standardize their energies and/or luminosities. Until now, a lot of GRB correlations have been proposed.
\begin{itemize}
\item
Amati correlation. The isotropic
energy $E_{\rm iso}$ of GRB prompt emission is associated with the rest-frame peak energy
of the prompt spectrum, i.e., $E_{\rm p}\propto E_{\rm iso}^{0.52}$ \cite{2002A&A...390...81A}.
\item
Yonetoku correlation. The correlation $L_{\rm iso}\propto E_{\rm
	p}^2$ is found with a sample of 16 GRBs
\citep{2004ApJ...609..935Y}.
\item
Ghirlanda correlation. A tight correlation between spectral peak energy $E_{\rm peak}$ and collimated energy $E_\gamma$ was
discovered \citep{2004ApJ...616..331G,2004ApJ...612L.101D} using 15 GRBs. The intrinsic scatter is up to $0.1$.
\item
Dainotti correlation. A tight
correlation between the X-ray afterglow parameters\citep{2008MNRAS.391L..79D}: $L_X$ and $T_a$, where $L_X$ is the luminosity of an X-ray plateau, and $T_a$ is
the time at which the X-ray light curve establishes a normal
power-law decay. The intrinsic scatter is $0.22$.
\item
Liang-Zhang correlation. Without imposing any theoretical model, an empirical correlation was found \citep{2005ApJ...633..611L} with 15 bursts among the isotropic energy of the prompt gamma-ray emission $E_{\rm iso}$, the
rest-frame peak energy $E_{\rm p}$, and the rest-frame break time in
the optical afterglow light curves $t_{\rm break}$.
\end{itemize}
According to the properties of the involved parameter, these correlations can be divided into three categories such as prompt correlations (i.e., the first three ones), afterglow correlations (i.e., the Dainotti one), and prompt-afterglow correlations (i.e., the Liang-Zhang one).

\begin{figure}
\begin{center}
	\includegraphics[width=\textwidth]{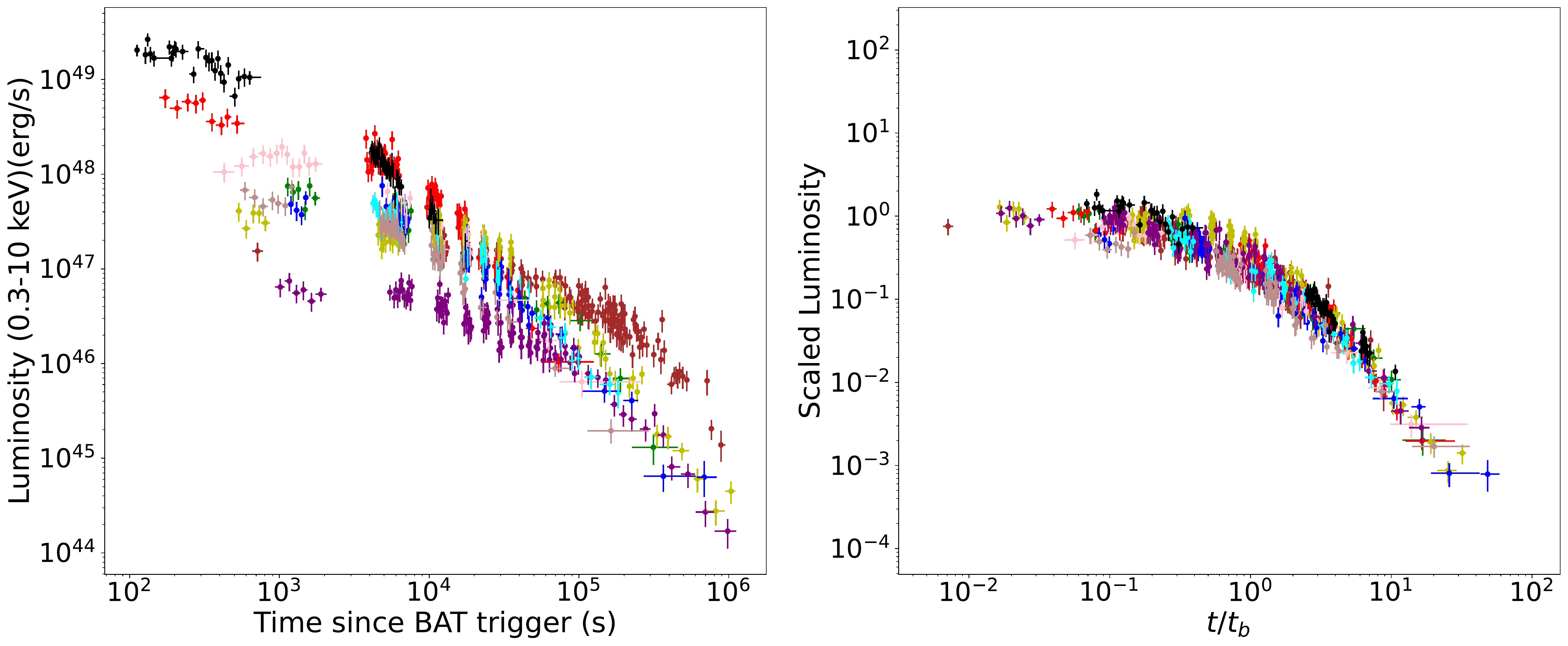}
	\caption{\label{G2} Standardized the light curves of LGRBs. Original and scaled plateau light curves of 10 LGRBs. Left panel shows the original X-ray (0.3-10 keV) light curves. Right panel shows the scaled light curves of the same GRBs. The dispersion of the scaled luminosity is 0.5 dex. Adapted from \cite{Wang2022}}
	\end{center}
\end{figure}

\subsection{Cosmological Constraints}

The most common method to constrain the cosmological parameters is by using the $\chi^2$ statistic. For an instance, in the standard cosmological model, the corresponding likelihood function can be written as
\begin{equation}
	\label{eq:chi}
	\chi^{2}(\Omega_{i})=\sum_{i=1}^{N}
	\frac{[\mu_{\rm th}(z,\Omega_{i})-\mu_{\rm obs}]^{2}}
	{\sigma_{\mu_{\rm obs}}^{2}},
\end{equation}
where $\Omega_{i}$ represents cosmological parameters, $\mu_{\rm th}(z,\Omega_{i})$ and $\mu_{\rm obs}$ the theoretical and observed distance modulus, respectively. While the observed distance modulus can be derived from GRB correlations, the theoretical one can be calculated by taking the follow equation
\begin{equation}
	\mu_{\rm th}=5\log(d_{\rm L})+25,
\end{equation}
where $d_{\rm L}$ is the luminosity distance. For a flat universe, the expressed form of luminosity distance can be written as
\begin{eqnarray}
	\label{eq:dL}
	d_{\rm L} = \frac{c(1+z)}{H_{0}} \int_{0}^{z} \frac{dz'}{\sqrt{\Omega_{\rm m} (1+z')^{3} + \Omega_{\Lambda}}},
\end{eqnarray}
where $\Omega_{\rm m}$, $\Omega_{\Lambda}$, $H_{0}$ and $c$ represent the cosmic matter density, dark energy density, Hubble constant and the speed of light, respectively. The best-fitting results of $\Omega_{\rm m}$, $\Omega_{\Lambda}$, $H_{0}$ can be given by minimizing Eq. (\ref{eq:chi}). For different cosmological models, Eq. (\ref{eq:dL}) needs to take corresponding modifications.

So far, a lot of effort has been made to constrain cosmological parameters
using GRBs since their cosmological origin was confirmed. However, due to the lack of low-redshift GRBs, the GRB correlations are cosmology-dependent. This is the so-called ``circularity problem". Many methods have been proposed to solve this problem, including fitting the cosmological parameters and luminosity correlation simultaneously, and calibrating GRB relation utilizing
other observations (i.e., SNe Ia). The first method is fitting the cosmological parameters and
luminosity correlation simultaneously \citep{2004ApJ...612L.101D}.
The second method is to calibrate GRB correlation with other observation data at low redshifts.
Recently, the light curves of LGRBs showing plateau phases in X-ray afterglows are standardized \cite{Wang2022}. The standardized result is shown in Fig. \ref{G2}. The Hubble parameter data is used to calibrate this correlation. The cosmological constraints are $\Omega_{\rm m} =0.32^{+0.05}_{-0.10}$ and $\Omega_{\Lambda}$ = $1.10^{+0.12}_{-0.31}$ (1$\sigma$) in the flat universe.
Third, the circularity problem could be partially solved by analyzing a sample of GRBs within a small redshift bin
\citep{Liang2006}. The advent of the multi-messenger era provides more solutions to the circularity problem. For example, the gravitational wave events with EM counterparts can be used to calibrate GRB correlations.

\acknowledgement This chapter is supported by the National Key R\&D Program of China (Grant No. 2021YFA0718500 and 2018YFA0404204), the China Manned Spaced Project (CMS-CSST-2021-A12), and the National Natural Science Foundation of China (Grant No. 11833003, U1831207, U2038105, and 12121003).

\bibliographystyle{aasjournal}

\end{document}